\documentclass[a4paper,11pt]{article}
\pdfoutput=1 

\usepackage{jheppub} 
\usepackage{amsthm,amsmath,amssymb}
\usepackage{mathrsfs}
\usepackage[citecolor=blue]{hyperref}
\usepackage{verbatim}
\usepackage{setspace}
\usepackage{array,multirow}
\usepackage[normalem]{ulem}
\usepackage{float}
\usepackage{rotating}
\usepackage{makecell}
\usepackage{mathdots}
\usepackage{color}
\usepackage[table]{xcolor}
\usepackage{adjustbox}
\usepackage{tikz}
\usepackage{pifont}
\usepackage[all]{xy}
\usepackage{bm}
\usepackage[T1]{fontenc} 
\usepackage{caption}
\usepackage{subcaption}

\def\fkg{\mathfrak{g}}

\def\CH{\mathcal{H}}
\def\CI{\mathcal{I}}
\def\CM{\mathcal{M}}
\def\CN{\mathcal{N}}
\def\CO{\mathcal{O}}
\def\CT{\mathcal{T}}

\def\bbC{\mathbb{C}}
\def\bbR{\mathbb{R}}
\def\bbZ{\mathbb{Z}}

\def\bfk{\mathbf{k}}
\def\bfl{\mathbf{l}}
\def\bfm{\mathbf{m}}
\def\bfn{\mathbf{n}}
\def\bfw{\mathbf{w}}
\def\bfy{\mathbf{y}}
\def\bfz{\mathbf{z}}

\def\tm{\tilde{m}}
\def\tq{\tilde{q}}

\def\tr{\mathrm{Tr}}

\title{Superconformal indices of $\mathcal{N}=4$ Chern-Simons matter theories }

\author[a]{Bohan Li}
\author[b,c]{Dan Xie}
\author[a]{and Wenbin Yan}

\affiliation[a]{Yau Mathematics Science center, Tsinghua University,  Beijing, 100084,China}
\affiliation[b]{Department of Mathematics, Tsinghua University,  Beijing, 100084,China}
\affiliation[c]{Peng Huanwu Center for Fundamental Theory, University of Science and Technology of China, Hefei, Anhui 230026, China}

\emailAdd{libh19@mails.tsinghua.edu.cn}
\emailAdd{danxie@mail.tsinghua.edu.cn}
\emailAdd{wbyan@mail.tsinghua.edu.cn}

\abstract{Gaiotto and Witten found that one can construct 3d $\mathcal{N}=4$ Chern-Simons matter theories by using $\mathcal{N}=4$ SCFT whose momentum map of global symmetries satisfy special condition. 
Usually, one uses free hypermultiplet and twisted hypermultiplet, and more recently it was found that strongly coupled theory such as 3d version of $T_N$ theory and Argyres-Douglas matter can also be used.
In this paper, we compute superconformal index of these $\mathcal{N}=4$ theories and derive the Coulomb/Higgs limit. Our results determine the moduli space of vacua, which is
used to check various interesting mirror symmetry involving CSM theory and usual $\mathcal{N}=4$ gauge theory.
}

\begin{document} 
\maketitle

\section{Introduction}

The supersymmetric index \cite{Bhattacharya_2008,Bhattacharya_2009,Kim:2009wb, Imamura:2011su, Kapustin:2011jm, Dimofte:2011py, Aharony:2013dha, Aharony:2013kma} is an invaluable tool to study the infrared (IR) behaviors of the three dimensional (3d) supersymmetric  theories. It counts the protected operators, and can be used to check various dualities, study the moduli spaces of vacua, etc. One particular application of the index is to study the possible enhanced symmetry in the IR, as one can extract various supersymmetric multiplets with their multiplicities from the index, and enhanced global symmetry or supersymmtry means corresponding multiplets should appear in the index.

We are interested in 3d $\CN=3$ Chern-Simons theories coupled with  $\CN=4$ SCFT $\CT$ with global symmetry group, and we are interested in the SUSY enhancement of those theories. Let the flavor symmetry of $\CT$ be $G_1\times G_2\times \ldots \times G_l$, and let $\mu_i$ be the moment map of $G_i$, it was argued in \cite{Gaiotto:2008sd, Assel:2022row} that if
\begin{equation}
\tr\mu_1^2=\tr\mu_2^2=\cdots=\tr\mu_l^2,
\label{moment}
\end{equation}
one can couple each $G_i$ with $\CN=3$ CS vector multiplet  at level $k_i$, and the resulting theory enhances to $\CN=4$ in the IR if the CS levels satisfy the following balance condition
\begin{equation}
\sum_{i=1}^l\frac{1}{k_i}=0.
\end{equation}
It was found in \cite{Gaiotto:2008sd} that bi-fundamental hypermultiplets or twisted hypermultiplets satisfy the momentum map relation \ref{moment}. 
Interestingly, many strongly coupled $\mathcal{N}=4$ SCFT derived from compactification of 4d $\mathcal{N}=2$ SCFT satisfy the momentum map relation. These 
theories include 3d $T_N$ theory \cite{Benini:2009mz}, and Argyres-Douglas matter \cite{Xie:2017vaf,Xie:2017aqx,Li:2023}, so naturally one get a large class of new $\mathcal{N}=4$ Chern-Simons matter (CSM) theories \cite{Assel:2022row,Li:2023}.

The purpose of this paper is to use the supersymmetric indices  to study various properties of these newly discovered $\CN=4$ CSM theories. The idea is to first compute the $\CN=2$ index  while keep track of the $U(1)$ R-symmetry which will be the $U(1)_{H-C}$ (R symmetry on the Higgs or Coulomb branch) after the enhancement. Because the index is invariant under the RG flow, this will also be the $\CN=4$ index of the IR theory.  Once the $\CN=4$ index is obtained, we can then take the Higgs/Coulomb limit \footnote{For CSM theories, there is no usual $\mathcal{N}=4$ Coulomb branch defined by turning on expectation value of scalar in $\mathcal{N}=4$ vector multiplet. However, the $R$ symmetry of a $\mathcal{N}=4$ SCFT is $SU(2)_A\times SU(2)_B$ and one can have two distinct type of branches in the moduli space depending on which $SU(2)_i$ symmetry acts on. We use the name Higgs/Coulomb 
to distinguish these two types of branches.}
 of the index to study the Higgs/Coulomb branches of the corresponding theory. One consistency check is that the Higgs/Coulomb branches of $\CN=4$ theories are hyper-Kahler manifolds, therefore our Higgs/Coulomb indices should reflect this (e.g. the complex dimension of the moduli space read from the index should be even).

We work out the indices for CSM theories with bi-fundamental matters, CSM theories of linear quivers, and CS theories coupled to multiple $T_2$ theories, and use the results to study their moduli spaces. 
\begin{enumerate}
\item We first compute the index for some abelian CSM theories, and found the Higgs/Coulomb limit. This provides verification of the duality proposal made in \cite{Jafferis:2008em}: a CSM theory is dual to a SCFT defined by usual 
$\mathcal{N}=4$ gauge theory. We also find interesting CSM theory whose Higgs branch and Coulomb branch are both the same, which deserves further studies. 
\item We then study non-Abelian $U(N_1)_k\times U(N_2)_{-k}$ CSM theories. Our result shows that the moduli space  receives quantum correction and we compute its exact Higgs branch index.
 We also generalize these results to non-Abelian CSM theories of a linear quiver, and provide a gluing formula for their Higgs/Coulomb indices.
These gluing formula helps us to find new mirrors between CSM theories and usual $\mathcal{N}=4$ gauge theory.

\item  Finally, we study the index of CSM theories coupled with $T_2$ theories. The Higgs branch is conjectured to be $\bbC^2/\hat{D}_{D+1}$, and $D$ is determined by the three CS levels.

\end{enumerate}


The paper is organized as follows. In section \ref{sec:AbelianCSM} we use Abelian CSM theories which enhances to $\CN=4$ as warm-up examples to illustrate the computation of the full indices and its implication on Higgs/Coulomb branches. Section \ref{sec:CSMlinearquiver} is devoted to Non-Abelian CSM theories of a linear quiver which enhances to $\CN=4$. Information on the Higgs and Coulomb branches can be extracted from indices, and we will see that the Higgs branches get quantum corrections and can be described by Coulomb branches of certain 3d SYM theories without CS terms. In section \ref{sec:indCBT2} we discuss indices and moduli spaces of  $T_2$ theories coupled to CS gauge theory. Reviews of 3d indices and some technical details are put in the appendix.

While this manuscript is prepared, we notice that another paper \cite{Comi:2023lfm} also discusses indices,  moduli spaces, and one-form symmetries of $T_N$ theories coupled to CS terms. Our results agree with each other and it would be nice to also apply techniques used in \cite{Comi:2023lfm} to study other CSM theories discussed here.

\section{Abelian  $\CN=4$ CSM theories}
\label{sec:AbelianCSM}

In this section we work out the  indices  of different abelian CSM theories which enhances to $\CN=4$ in the IR, and their Higgs/Coulomb limits. The full indices provide good checks for mirror pairs between CSM theories and SYM theories discussed in \cite{Jafferis:2008em}. One can also read off the Higgs/Coulomb moduli spaces from the index.

\subsection{$U(1)_k\times U(1)_{-k}$ CSM theories}

\begin{figure}
\begin{center}

\tikzset{every picture/.style={line width=0.75pt}} 

\begin{tikzpicture}[x=0.75pt,y=0.75pt,yscale=-1,xscale=1]

\draw   (10,45) .. controls (10,25.67) and (25.67,10) .. (45,10) .. controls (64.33,10) and (80,25.67) .. (80,45) .. controls (80,64.33) and (64.33,80) .. (45,80) .. controls (25.67,80) and (10,64.33) .. (10,45) -- cycle ;
\draw   (130,45) .. controls (130,25.67) and (145.67,10) .. (165,10) .. controls (184.33,10) and (200,25.67) .. (200,45) .. controls (200,64.33) and (184.33,80) .. (165,80) .. controls (145.67,80) and (130,64.33) .. (130,45) -- cycle ;
\draw    (80,46) -- (130,46) ;

\draw (25,35) node [anchor=north west][inner sep=0.75pt]    {$U( 1)_{k}$};
\draw (140,35) node [anchor=north west][inner sep=0.75pt]    {$U(1)_{-k}$};
\draw (86,20) node [anchor=north west][inner sep=0.75pt]    {$X,\tilde{X}$};

\end{tikzpicture}

\end{center}
\caption{\label{fig:CSMU1U2}$U(1)_k\times U(1)_{-k}$ $\CN=4$ CSM theory with one bifundamental hypermultiplet $(X,\tilde{X})$.}
\end{figure}
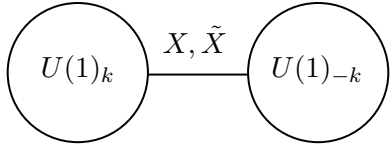

The simplest example is the $U(1)_k\times U(1)_{-k}$ CSM theories with an $\CN=4$ bi-fundamental hypermultiplet as in figure \ref{fig:CSMU1U2}.
Its  index (see appendix \ref{sec:3dindex} for the details of 3d $\mathcal{N}=2$ index) is
\begin{equation}
\label{eq:indexU1U1}
\begin{split}
\CI^{U(1)_k\times U(1)_{-k}}=&
\sum_{m_i\in\bbZ}\int \prod_{i=1}^2\frac{dz_i}{2\pi iz_i} w_i^{m_i}z_i^{k_im_i}
 Z_{hyp}(\{z_1,m_1\},\{z_2,m_2\};t)\\
 =&\sum_{m_i\in\bbZ}\int \prod_{i=1}^2\frac{dz_i}{2\pi iz_i}  w_i^{m_i}z_i^{k_im_i}
 \left(\frac{x}{t^2}\right)^{\frac{1}{2}|m_1-m_2|}
 \frac{((-1)^{m_1-m_2}t^{-1}x^{\frac{3}{2}+|m_1-m_2|}(z_1/z_2)^\pm ;x^2)}{((-1)^{m_1-m_2}tx^{\frac{1}{2}+|m_1-m_2|}(z_1/z_2)^\pm ;x^2)}
\end{split}
\end{equation}
where $k_1=-k_2=k>0$ and $(a z^\pm;x^2)\equiv(a z;x^2)(a z^{-1};x^2)$. Notice that if we change the gauge fugacities to $u=z_1/z_2$ and $v=z_1z_2$, integration of $v$ imposes the constraint $m_1=m_2$, so only monopoles with charge $(m,m)$ contribute to the index.

Usually it is more convenient to write the result in terms of the single letter index rather than the full index. Define the normalized single letter index $\tilde{\CI}$ by
\begin{equation}
\tilde{\CI}=(1-x^2)\mathrm{PLog}[\CI],\quad \mathrm{i.e.}\quad\CI=\mathrm{PE}\left[\frac{\tilde{\CI}}{1-x^2}\right],
\end{equation}
where 
\begin{equation}
\mathrm{PLog}[f(x,y,z,\cdots)]=\sum_{n=1}^\infty\frac{\mu(n)}{n}f(x^n,y^n,z^n,\cdots).
\end{equation}
Here $\mu(n)$ is the Moebius $\mu$ function. The the normalized single letter indices for the first few values of $k$ are
\begin{equation}
\begin{split}
\tilde{\CI}^{U(1)_1\times U(1)_{-1}} = &
t(w+w^{-1})x^{\frac{1}{2}}-t^{-1}(w+w^{-1})x^{\frac{3}{2}}+\CO(x^{11}),\\
\tilde{\CI}^{U(1)_2\times U(1)_{-2}}
=&t^2(1+w+w^{-1})x -(t^2+(w+2+w^{-1}))x^2 +(t^2(w+3+w^{-1})+t^{-2})x^3\\
&-(t^4(2w+3+2w^{-1})+(3w+4+3w^{-1}))x^4+\cdots,\\
\tilde{\CI}^{U(1)_3\times U(1)_{-3}}
=&t^2x +t^3(w+w^{-1})x^{\frac{3}{2}} -2x^2-t(w+w^{-1})x^{\frac{5}{2}}+(-t^6+t^2+t^{-2})x^3\\
 &+t^3(w+w^{-1})x^{\frac{7}{2}}+(2t^4-3)x^4+\cdots,
\end{split}
\end{equation}
where $w=w_1w_2$ and the overall topological $U(1)$ is decoupled. From the index one can see that $U(1)_1\times U(1)_{-1}$ CS theory with hypermultiplet is the same as a free hypermultiplet.

\subsubsection{Higgs and Coulomb branch}

Unlike the full index, the Higgs and Coulomb index have closed form expressions. First rewrite the index \ref{eq:indexU1U1} in terms of $q$ and $\tilde{q}$
\begin{equation}
\CI^{U(1)_k\times U(1)_{-k}}=\sum_{m\in\bbZ}\int\frac{du}{2\pi i u}w^{m}u^{km}
\frac{(q^{\frac{1}{2}}\tilde{q} u^{\pm};q\tilde{q})}{(q^{\frac{1}{2}} u^\pm; q\tilde{q})}.
\end{equation}
It is obvious that the Coulomb limit of the index ($q\rightarrow 0$) is 1. On the other hand, the Higgs limit of the index is non-trivial
\begin{equation}
\begin{split}
\CI^{U(1)_k\times U(1)_{-k}}_H=\sum_{m\in\bbZ}w^m\int\frac{du}{2\pi i u}u^{km}\frac{1}{(1-q^{\frac{1}{2}}u)(1-q^{\frac{1}{2}}u^{-1})},
\end{split}
\end{equation}
with $w=w_1w_2$.
Define $g_1(q;m)$ as
\begin{equation}
\label{eq:defg1}
g_1(q;m)=\int\frac{du}{2\pi i u}u^{m}\frac{1}{(1-q^{\frac{1}{2}}u)(1-q^{\frac{1}{2}}u^{-1})}=\frac{q^{\frac{|m|}{2}}}{1-q},
\end{equation}
then the Higgs index is
\begin{equation}
\CI^{U(1)_k\times U(1)_{-k}}_H=\sum_{m\in\bbZ}w^m g_1(q,km)=\frac{1-q^k}{(1-q)(1-q^{\frac{k}{2}}w)(1-q^{\frac{k}{2}}w^{-1})},
\end{equation}
which is the same as the Hilbert series of the coordinate ring of $\bbC^2/\bbZ_k$. We deduce that the Higgs branch of the $U(1)_k\times U(1)_{-k}$ CSM theory is $\bbC^2/\bbZ_k$ from the index. 
The ideal defining the coordinate ring of $\bbC^2/\bbZ_k$ is $uv+z^k=0$, and its Hilbert series is just $\frac{1-q^k}{(1-q)(1-q^{\frac{k}{2}}w)(1-q^{\frac{k}{2}}w^{-1})}$, here the charges of the coordinates 
are $u=(\frac{k}{2},1), v=(\frac{k}{2},-1), z=(1,0)$.

\subsection{$U(1)_k\times U(1)_{-k}\times U(1)_k$ CSM theories}
\label{sec:U1U1U1}

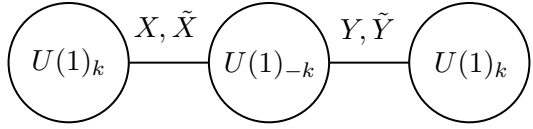
\begin{figure}
\begin{center}

\tikzset{every picture/.style={line width=0.75pt}} 

\begin{tikzpicture}[x=0.75pt,y=0.75pt,yscale=-1,xscale=1]

\draw   (10,40) .. controls (10,23.43) and (23.43,10) .. (40,10) .. controls (56.57,10) and (70,23.43) .. (70,40) .. controls (70,56.57) and (56.57,70) .. (40,70) .. controls (23.43,70) and (10,56.57) .. (10,40) -- cycle ;
\draw   (110,40) .. controls (110,23.43) and (123.43,10) .. (140,10) .. controls (156.57,10) and (170,23.43) .. (170,40) .. controls (170,56.57) and (156.57,70) .. (140,70) .. controls (123.43,70) and (110,56.57) .. (110,40) -- cycle ;
\draw   (210,40) .. controls (210,23.43) and (223.43,10) .. (240,10) .. controls (256.57,10) and (270,23.43) .. (270,40) .. controls (270,56.57) and (256.57,70) .. (240,70) .. controls (223.43,70) and (210,56.57) .. (210,40) -- cycle ;
\draw    (70,40) -- (110,40) ;
\draw    (170,40) -- (210,40) ;

\draw (20,31.4) node [anchor=north west][inner sep=0.75pt]    {$U( 1)_{k}$};
\draw (221,32.4) node [anchor=north west][inner sep=0.75pt]    {$U( 1)_{k}$};
\draw (116,32.4) node [anchor=north west][inner sep=0.75pt]    {$U( 1)_{-k}$};
\draw (71,12.4) node [anchor=north west][inner sep=0.75pt]    {$X,\tilde{X}$};
\draw (174,13.4) node [anchor=north west][inner sep=0.75pt]    {$Y,\tilde{Y}$};

\end{tikzpicture}

\end{center}
\caption{\label{fig:threeU1CSM}The quiver diagram of the $U(1)_k\times U(1)_{-k}\times U(1)_k$ CSM theory. $(X,\tilde{X})$ is the hypermultiplet and $(Y,\tilde{Y})$ is the twisted hypermultiplet.}
\end{figure}

Now we consider the index of $\CN=3$ $U(1)_k\times U(1)_{-k}\times U(1)_k$ CSM theories with a hypermultiplet of $U(1)^3$ charges $(1,-1,0)$ and a twisted hypermultiplet of $U(1)^3$ charges $(0,-1,1)$ (figure \ref{fig:threeU1CSM}) which enhances to $\CN=4$ in the IR \cite{Jafferis:2008em}.  Its superconformal index is
\begin{equation}
\begin{split}
\CI^{U(1)^3}=
\sum_{m_i\in\bbZ}\int &\prod_{i=1}^3\frac{dz_i}{2\pi iz_i} \prod_{i=1}^3 w_i^{m_i}z_i^{k_im_i}Z_{vec}(\{z_i,m_i\}) \\
&\times Z_{hyp}(\{z_1,m_1\},\{z_2,m_2\};t)Z_{hyp}(\{z_2,m_2\},\{z_3,m_3\};t^{-1}),
\end{split}
\end{equation}
with $k_1=-k_2=k_3=k>0$.

When $k=1$, the superconformal index can be worked out as a power series of $x$,
\begin{equation}
\begin{split}
\CI^{U(1)^3,k=1}
=&1+\left(t^2\left(w_1w_2+1+\frac{1}{w_1w_2}\right) + t^{-2}\left(w_2w_3+1+\frac{1}{w_2w_3}\right) \right) x\\
&+\left(t^4\left(\sum_{i=-2}^2w_1^iw_2^i\right) - \left(w_1w_2+w_2w_3+3+\frac{1}{w_1w_2}+\frac{1}{w_2w_3}\right) +t^{-4}\left(\sum_{i=-2}^2w_2^iw_3^i\right)\right) x^2 +\cdots \\
\xrightarrow{w_i\rightarrow 0} & 1+ (3t^2+3t^{-2})x +(5t^4-7+5t^{-4})x^2 +(7t^6-4t^2-4t^{-2}+7t^6)x^3\\
&+(9t^8-4t^4+7-4t^{-4}+9t^{-8})x^4+(11t^{10}-4t^6-20t^2-20t^{-2}-4t^{-6}+11t^{-1)})x^5+\cdots.
\end{split}
\end{equation}
Out of three $w_i$'s only combination $w_1w_2$ and $w_2w_3$ appear in the index, since the overall topological $U(1)$ is decoupled.
Its Higgs and Coulomb branch indices have the following form,
\begin{equation}
\begin{split}
\CI^{U(1)^3,k=1}_H &=\frac{1-q^2}{(1-q)(1-q w_1w_2)(1-q w_1^{-1}w_2^{-1})},\\
\CI^{U(1)^3,k=1}_C &=\frac{1-\tilde{q}^2}{(1-\tilde{q})(1-\tilde{q} w_2w_3)(1-\tilde{q } w_2^{-1}w_3^{-1})}.
\end{split}
\end{equation}
These results confirm that both the Higgs and Coulomb branches of $U(1)_1\times U(1)_{-1}\times U(1)_1$ CSM theory are $\bbC^2/\bbZ_2$  \cite{Jafferis:2008em}. Moreover, this theory is self-mirror and is proposed to dual to the $\CN=4$ QED with two flavors  \cite{Jafferis:2008em}. Both the Higgs 
branch and Coulomb branch of this theory is $\bbC^2/\bbZ_2$. We have checked that its full  index is indeed the same as the  index of the $\CN=4$ QED with two flavors order by order.

For generic integer level $k$, one finds that the Higgs and Coulomb limit of the index is
\begin{equation}
\begin{split}
\CI^{U(1)^3,k}_H &=\frac{ 1-q^{k+1} }{(1-q)(1- q^{\frac{k+1}{2}} w_1w_2)(1- q^{\frac{k+1}{2}} w_1^{-1}w_2^{-1})},\\
\CI^{U(1)^3,k}_C &=\frac{ 1-\tilde{q}^{k+1} }{(1-\tilde{q})(1- \tilde{q}^{\frac{k+1}{2}} w_2w_3)(1- \tilde{q}^{\frac{k+1}{2}} w_2^{-1}w_3^{-1})}.
\end{split}
\end{equation}
The indices confirms that the $U(1)_k\times U(1)_{-k}\times U(1)_k$ CSM theory is again self-mirror with both Higgs and Coulomb branch being $\bbC^2/\bbZ_{k+1}$ \cite{Assel:2017eun}. 
We are not aware of usual $\mathcal{N}=4$ theories with such self-dual moduli space, and it would be interesting to further study them.

\subsection{$U(1)_1\times U(1)_{-1}\times U(1)_1\times U(1)_{-1}$ CSM theory}
\label{sec:U1U1U1U1}

\begin{figure}
 \begin{center}

\tikzset{every picture/.style={line width=0.75pt}} 

\begin{tikzpicture}[x=0.75pt,y=0.75pt,yscale=-1,xscale=1]

\draw   (10,40) .. controls (10,23.43) and (23.43,10) .. (40,10) .. controls (56.57,10) and (70,23.43) .. (70,40) .. controls (70,56.57) and (56.57,70) .. (40,70) .. controls (23.43,70) and (10,56.57) .. (10,40) -- cycle ;
\draw   (110,40) .. controls (110,23.43) and (123.43,10) .. (140,10) .. controls (156.57,10) and (170,23.43) .. (170,40) .. controls (170,56.57) and (156.57,70) .. (140,70) .. controls (123.43,70) and (110,56.57) .. (110,40) -- cycle ;
\draw   (210,40) .. controls (210,23.43) and (223.43,10) .. (240,10) .. controls (256.57,10) and (270,23.43) .. (270,40) .. controls (270,56.57) and (256.57,70) .. (240,70) .. controls (223.43,70) and (210,56.57) .. (210,40) -- cycle ;
\draw    (70,40) -- (110,40) ;
\draw    (170,40) -- (210,40) ;
\draw   (310,40) .. controls (310,23.43) and (323.43,10) .. (340,10) .. controls (356.57,10) and (370,23.43) .. (370,40) .. controls (370,56.57) and (356.57,70) .. (340,70) .. controls (323.43,70) and (310,56.57) .. (310,40) -- cycle ;
\draw    (270,40) -- (310,40) ;

\draw (21,32.4) node [anchor=north west][inner sep=0.75pt]    {$U( 1)_{k}$};
\draw (221,32.4) node [anchor=north west][inner sep=0.75pt]    {$U( 1)_{k}$};
\draw (116,32.4) node [anchor=north west][inner sep=0.75pt]    {$U( 1)_{-k}$};
\draw (71,12.4) node [anchor=north west][inner sep=0.75pt]    {$X,\tilde{X}$};
\draw (174,13.4) node [anchor=north west][inner sep=0.75pt]    {$Y,\tilde{Y}$};
\draw (321,32.4) node [anchor=north west][inner sep=0.75pt]    {$U( 1)_{-k}$};
\draw (274,13.4) node [anchor=north west][inner sep=0.75pt]    {$Z,\tilde{Z}$};

\end{tikzpicture}

 \end{center}
 \caption{\label{fig:fourU1CSM}The quiver diagram of the $U(1)_k\times U(1)_{-k}\times U(1)_k\times U(1)_{-k}$ CSM theory. Here $(X,\tilde{X})$ and $(Z,\tilde{Z})$ are twisted hypermultiplets, and $(Y,\tilde{Y})$ is a hypermultiplet.}
 \end{figure}
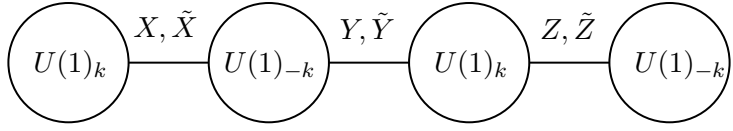

As a final example of Abelian CSM theories we consider the index of $\CN=3$ $U(1)_k\times U(1)_{-k}\times U(1)_k\times U(1)_{-k}$ CSM theories with two twisted hypermultiplets of gauge charges $(1,-1,0,0)$ and $(0,0,-1,1)$ and a  hypermutliplet of gauge charges $(0,-1,1,0)$ (figure \ref{fig:fourU1CSM}). This theory enhances to $\CN=4$ in the IR \cite{Jafferis:2008em}, and its superconformal index is
\begin{equation}
\begin{split}
\CI^{U(1)^4}=
&\sum_{m_i\in\bbZ}\int \prod_{i=1}^4\frac{dz_i}{2\pi iz_i} \prod_{i=1}^4 w_i^{m_i}z_i^{k_im_i}Z_{vec}(\{z_i,m_i\}) \\
&\times Z_{hyp}(\{z_1,m_1\},\{z_2,m_2\};t^{-1})Z_{hyp}(\{z_2,m_2\},\{m_3,z_3\};t) Z_{hyp}(\{z_3,m_3\},\{z_4,m_4\};t^{-1}),
\end{split}
\end{equation}
with $k_1=-k_2=k_3=-k_4=k>0$.

When $k=1$, the (unrefined) full index is
\begin{equation}
\begin{split}
\CI^{U(1)^4,k=1}\xrightarrow{w_i\rightarrow 1}&1+(t^2+8t^{-2})x+2t^3x^{\frac{3}{2}}+(t^4-10+27t^{-4})x^2 \\ 
&+(2t^5-2t)x^{\frac{5}{2}} +(3t^6 -46t^{-2}+64 t^{-6})x^3+\cdots.
\end{split}
\end{equation} 
Its Higgs index is simple
\begin{equation}
\CI^{U(1)^4,k=1}_H=\frac{ 1-q^3 }{ (1-q) (1-q^{\frac{3}{2}} w_2w_3) (1-q^{\frac{3}{2}} w^{-1}_2w^{-1}_3)}.
\end{equation}
This results confirms that the Higgs branch of this $U(1)^4$ CSM is $\bbC^2/\bbZ_3$, which is the same as the Coulomb branch of $\CN=4$ QED with $N_f=3$  \cite{Jafferis:2008em}. The Coulomb index is the infinite series
\begin{equation}
\begin{split}
\CI^{U(1)^4,k=1}_C
&=\sum_{n=0}^\infty\tilde{q}^n\chi_{[n,n]}^{SU(3)}=1+\tilde{q}\chi_{\mathbf{8}}^{SU(3)}+\tilde{q}^2\chi_{\mathbf{27}}^{SU(3)}+\tilde{q}^3\chi_{\mathbf{64}}^{SU(3)}+\cdots \\
&=\mathrm{PE}\left [
\tilde{q}\chi_{\mathbf{8}}^{SU(3)}-\tilde{q}^2(1+\chi_{\mathbf{8}}^{SU(3)})+2\tilde{q}^3\chi_{\mathbf{8}}^{SU(3)}+\cdots\right],
\end{split}
\end{equation}
where $\chi_{R}^{SU(3)}$ is the character of the representation $R$ of $SU(3)$. One can check order by order that $\CI^{U(1)^4,k=1}_C$ is the same as the Higgs index of $\CN=4$ QED with $N_f=3$, therefore the Coulomb branch of this $U(1)^4$ CSM at $k=1$ is the same as the Higgs branch of $\CN=4$ QED with $N_f=3$ which is the hyperkahler quotient
\begin{equation}
\left\{ \sum_{i=1}^3 (|Q_i|^2-|\tilde{Q}^i|^2)=0,\ \sum_{i=1}^3Q_i\tilde{Q}^i=0\right\} / U(1),
\end{equation}
which indeed has an isometry of $SU(3)$.
One can also see that full indices of thest two theories indeed exchange to each other under the transformation $t\mapsto t^{-1}$, which provides further evidence that the $U(1)_1\times U(1)_{-1}\times U(1)_1\times U(1)_{-1}$ CSM theory is mirror to the $\CN=4$ QED with three flavors  \cite{Jafferis:2008em}.

\section{Non-Abelian $\CN=4$ CSM theories }
\label{sec:CSMlinearquiver}

Now we study the indices of non-Abelian CSM theories which enhance to $\CN=4$ in the IR, and their Higgs/Coulomb limit. We will see that in many examples their Higgs branches receives quantum correction and are the same as Coulomb branches of certain SYM theories without CS terms. 

\subsection{$\CN=4$ CSM theories with bi-fundamentals}

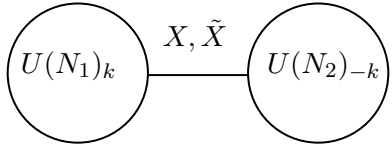
\begin{figure}
\begin{center}

\tikzset{every picture/.style={line width=0.75pt}} 

\begin{tikzpicture}[x=0.75pt,y=0.75pt,yscale=-1,xscale=1]

\draw   (10,45) .. controls (10,25.67) and (25.67,10) .. (45,10) .. controls (64.33,10) and (80,25.67) .. (80,45) .. controls (80,64.33) and (64.33,80) .. (45,80) .. controls (25.67,80) and (10,64.33) .. (10,45) -- cycle ;
\draw   (130,45) .. controls (130,25.67) and (145.67,10) .. (165,10) .. controls (184.33,10) and (200,25.67) .. (200,45) .. controls (200,64.33) and (184.33,80) .. (165,80) .. controls (145.67,80) and (130,64.33) .. (130,45) -- cycle ;
\draw    (80,46) -- (130,46) ;

\draw (15,32.4) node [anchor=north west][inner sep=0.75pt]    {$U( N_{1})_{k}$};
\draw (138,32.4) node [anchor=north west][inner sep=0.75pt]    {$U( N_{2})_{-k}$};
\draw (86,16.4) node [anchor=north west][inner sep=0.75pt]    {$X,\tilde{X}$};

\end{tikzpicture}

\end{center}
\caption{\label{fig:CSMUn1Un2}$\CN=4$ CSM theory with one bifundamental hypermultiplet.}
\end{figure}

First let us look at the indices of non-Abelian CSM theories with gauge group $U(N_1)_{k_1}\times U(N_2)_{k_2}$ and a  bi-fundamental hypermutiplet (figure \ref{fig:CSMUn1Un2}). These theories enhance to $\CN=4$ if $k_1=-k_2=k>0$ \cite{Gaiotto:2008sd}. The $\CN=2$ indices of these theories were discussed and used to check some dualities in \cite{Nosaka:2018eip}. Using the $\CN=4$ indices, we also confirm these duality; The Higgs and Coulomb branches of these theories are also discussed.

This  $U(N_1)_k\times U(N_2)_{-k}$ CS theory is good or ugly in the sense of \cite{Gaiotto:2008ak} when $k\geq N_1+N_2-1$ \cite{Nosaka:2018eip}, and we compute indices only for non-bad theories. The result is:
\begin{equation}
\begin{split}
\label{eq:UN1UN2full}
&\CI^{U(N_1)_k\times U(N_2)_{-k}} \\
=&\sum_{m_1\geq m_2\geq\cdots \geq m_{N_1}}\sum_{n_1\geq n_2\geq\cdots\geq n_{N_2}}\frac{1}{W(\bfm)W(\bfn)}\int \frac{d\bfy}{2\pi i\bfy}\frac{d\bfz}{2\pi i \bfz}\left(\prod_{i=1}^{N_1}u^{m_i}y_i^{km_i}\right)
\left(\prod_{i=1}^{N_2}v^{-n_i}z_i^{-kn_i}\right)\\
&\times Z_{vec}(\{\bfy,\bfm\}) Z_{hyp}(\{\bfy,\bfm\},\{\bfz,\bfn\}) Z_{vec}(\{\bfz,\bfn\}).
\end{split}
\end{equation}
Here $u$ and $v$ are fugacities of the two $U(1)$ topological symmetries. Similar to the $U(1)$ case, integrating over the decoupled gauge $U(1)$ imposes the constraint
\begin{equation}
\sum_im_i=\sum_jn_j,
\end{equation}
which means that one topological $U(1)$ is decoupled, only the combination $w=u/v$ appear in the final index.

The index of the monopole operator $V_{\bfm,\bfn}$ can be read-off directly from \ref{eq:UN1UN2full} as
\begin{equation}
\label{eq:monopoleCharges}
w^{\sum_im_i}\left(\prod_i y^{km_i}_i \prod_j z_j^{-kn_j}\right)
q^{-\frac{1}{2}\sum_{i<j}|m_i-m_j|-\frac{1}{2}\sum_{i<j}|n_i-n_j| }
\tilde{q}^{ \frac{1}{2}\sum_{i,j}|m_i-n_j|-\frac{1}{2}\sum_{i<j}|m_i-m_j|-\frac{1}{2}\sum_{i<j} |n_i-n_j|  },
\end{equation}
which encodes the quantum corrected values of $\Delta-R_C$, $\Delta-R_H$, gauge charges and etc of the monopole operators $V_{\bfm,\bfn}$. This information will be crucial when taking Higgs/Coulomb limit.

\subsubsection{Examples of full indices of $U(N_1)_k\times U(N_2)_{-k}$ CSM theories}
\label{sec:exUN1UN2}

First consider $N_1=2$ and $N_2=1$, and the theory is non-bad when $k\geq2$. When $k=2$ the full index is 
\begin{equation}
\CI^{U(2)_2\times U(1)_{-2}}=\mathrm{PE}
\left[
\frac{t x^{\frac{1}{2}}-t^{-1}x^{\frac{3}{2}}}{1-x^2}(w+w^{-1})
\right]+\CO(x^{11}),
\end{equation}
which is the same as the index of a free hypermultiplet with no background flux up to $\CO(x^{11})$
\begin{equation}
\CI_{1-hyp}=\frac{ (wt^{-1}x^{\frac{3}{2}};x^2)_\infty(w^{-1}t^{-1}x^{\frac{3}{2}};x^2)_\infty}{ (wt x^{\frac{1}{2}};x^2)_\infty(w^{-1}t x^{\frac{1}{2}};x^2)_\infty }.
\end{equation}
This result further confirms the proposal that $U(2)_2\times U(1)_{-1}$ CSM theory is dual to a free hypermultiplet \cite{Nosaka:2018eip}. The Higgs index for this theory is also the Hilbert series of $\bbC^2$.

When $k>2$, the full index lacks a closed form expression as in $k=2$ case.
For example when $k=3$ and $4$, the normalized single letter indices are
\begin{equation}
\begin{split}
\tilde{\CI}^{ U(2)_3\times U(1)_{-3} }
=&t^2(1+w+w^{-1})x-(t^4+2+w+w^{-1})x^2\\
&+(t^2(3+w+w^{-1})+t^{-2})x^3+\cdots,\\
\tilde{\CI}^{ U(2)_4\times U(1)_{-4} }
=&t^2x+t^3(w+w^{-1})x^{\frac{3}{2}}-2x^2-t(w+w^{-1})x^{\frac{5}{2}}\\
&+(-t^6+t^2+t^{-2})x^3+\cdots.
\end{split}
\end{equation}
%
%
%

When $N_1=N_2=2$, the theory is not bad if $k\geq 3$. For the first few values of $k$, the normalized single letter index is
\begin{equation}
\begin{split}
\tilde{\CI}^{ U(2)_3\times U(2)_{-3} }
=&t(w+w^{-1})x^{\frac{1}{2}} + t^2x +(t^3-t^{-1})(w+w^{-1})x^{\frac{3}{2}}-2x^2 \\
&-t(w+w^{-1})x^{\frac{5}{2}}+(-t^6+t^2+t^{-2})x^3+\cdots,\\
\tilde{\CI}^{ U(2)_4\times U(2)_{-4} }
=& t^2(1+w+w^{-1})x + \left( t^4(1+w+w^{-1}) -(2+w+w^{-1}) \right)x^2 \\
& + \left( -t^6 - t^2(1+w+w^{-1}) +t^{-2} \right) x^3+ \left( -t^8 + t^4(3+w+w^{-1}) \right)x^4\cdots,\\
\tilde{\CI}^{ U(2)_5\times U(2)_{-5} }
=& t^2 x+t^3(w+w^{-1})x^{\frac{3}{2}} +(t^4-2)x^2+(t^5-t)(w+w^{-1})x^{\frac{5}{2}}+(-t^2+t^{-2})x^3\\
& -t^3(w+w^{-1})x^{\frac{7}{2}}+(-t^8+t^4)x^4+t^5(w+w^{-1})+(-t^{10}+2t^6-3t^2-t^{-2})x^5+\cdots,
\end{split}
\end{equation}
In all these examples, one can take the Higgs/Coulomb limit of the full index and in next sections we will provide closed form expressions for Higgs/Coulomb indices.


We also checked that the ratio of the $\CN=4$ index of the $U(2)_3\times U(2)_{-3}$ CSM theory and that of the $U(1)_3\times U(1)_{-3}$ theory is  the index of one hypermultiplet up to $\CO(x^6)$
\begin{equation}
\frac{\CI^{U(2)_3\times U(2)_{-3}}}{\CI^{U(1)_3\times U(1)_{-3}}} = \CI_{1-hyp} \Leftrightarrow 
\tilde{\CI}^{U(2)_3\times U(2)_{-3}} - \tilde{\CI}^{U(1)_3\times U(1)_{-3}}
=\left(tx^{\frac{1}{2}}-t^{-1}x^{\frac{3}{2}}\right)(w+w^{-1})+\CO(x^6),
\end{equation}
which confirms the duality of \cite{Nosaka:2018eip} using the $\CN=4$ index.

\subsubsection{The Higgs index}

Let us  work out the closed form expression of the Higgs index. Firstly consider the case $N_1=N_2=N$,  the index of the monopole operator $V_{\bfm,\bfn}$ is
\begin{equation}
w^{\sum_im_i}\left(\prod_i y^{km_i}_i z_i^{-kn_i}\right)
q^{-\frac{1}{2}\sum_{i<j}(|m_i-m_j|+|n_i-n_j|) }
\tilde{q}^{ \frac{1}{2}\sum_{i,j}|m_i-n_j|-\frac{1}{2}\sum_{i<j}(|m_i-m_j|+|n_i-n_j|)  }.
\end{equation}
Using the inequality $|m_i-n_j|+|n_j-m_k|\geq |m_i-m_k|$, we see that the $\Delta-R_H$ of $V_{\bfm,\bfn}$ (power of $\tilde{q}$) is always non-negative
\begin{equation}
\label{eq:DeltaRHcharge}
\sum_{1\leq i,j \leq N} |m_i-n_j|-\sum_{1\leq i<j \leq N}(|m_i-m_j|+|n_i-n_j|)\geq0,
\end{equation}
therefore we can safely taking the $\tilde{q}\rightarrow 0$ limit before the integration over gauge fugacities $\bfy$ and $\bfz$ in \ref{eq:UN1UN2full}. We find out that the Higgs index receives contribution only from monopoles $V_{\bfm, \bfm}$, because only when $\bfm=\bfn$ the monopole operator is in the Higgs branch and massless bifundamentals exist. Denoting $\bfm=(m_1,m_2,\cdots,m_N)$ by
\begin{equation}
\bfm=(s_1^{l_1}s_2^{l_2}\cdots s_r^{l_r}),\quad l_1+l_2+\cdots+l_r=N,
\end{equation}
with $s_1>s_2>\cdots>s_r$, the Higgs index with a background monopole charge $(\bfm,\bfm)$ is
\begin{equation}
\label{eq:bifundkm}
\CI^{U(N)_k\times U(N)_{-k},\bfm}_H=w^{\sum_im_i}q^{-\sum_{i<j}|m_i-m_j| }\prod_{i=1}^r g_{l_i}(q; kl_i|s_i|),
\end{equation}
where
\begin{equation}
g_l(q;m)=\frac{q^{\frac{|m|}{2}}}{\prod_{i=1}^l(1-q^i)}
\end{equation}
is  to the $l$-th symmetric power of $g_1(q;m)$ defined in \ref{eq:defg1},
\begin{equation}
g_l(q;lm)=\mathrm{Sym}^lg_1(q;m)=\mathrm{Res}_{\nu=0}\frac{1}{\nu^{l-1}}\exp\left(\sum_{n>0}\frac{\nu^n}{n}g_1(q^n;m) \right).
\end{equation}
Notice that the condition $k-2N\geq -1$ for the theory to be non-bad ensures that there is no negative powers of $q$ in $\CI^{U(N)_k\times U(N)_{-k},\bfm}_H$.

The Higgs index is then the sum of $\CI^{U(N)_k\times U(N)_{-k},\bfm}_H$'s over all possible monopole charges,
\begin{equation}
\label{eq:HiggsIndUNUNk}
\begin{split}
\CI^{U(N)_k\times U(N)_{-k}}_H=&\sum_{m_1\geq m_2\geq\cdots m_N}\CI^{U(N)_k\times U(N)_{-k},\bfm}_H\\
=&\sum_{m_1\geq m_2\geq\cdots m_N}w^{\sum_im_i}q^{-\sum_{i<j}|m_i-m_j| }\prod_{i=1}^r g_{l_i}(q; kl_i|s_i|).
\end{split}
\end{equation}

Surprisingly, after summing over $\bfm$'s, the result has a simple form
\begin{equation}
\CI^{U(N)_k\times U(N)_{-k}}_H=\prod_{i=1}^N\frac{1-q^{k+1-i}}{(1-q^i)(1-q^{\frac{k}{2}+1-i}w)(1-q^{\frac{k}{2}+1-i}w^{-1})}.
\end{equation}
The classical Higgs branch of $U(N)_k\times U(N)_{-k}$ CS matter theory is argued to be the $N$-th symmetric power of $\bbC^2/Z_k$ ($\mathrm{Sym}^N\bbC^2/Z_k$) \cite{Nosaka:2018eip}, however, the Higgs index is not the Hilbert series of the coordinate ring of $\mathrm{Sym}^N\bbC^2/Z_k$ because
\begin{equation}
\mathrm{Sym}^N\CI^{U(1)_k\times U(1)_{-k}}_H=\sum_{m_1\geq m_2\geq\cdots m_N}w^{\sum_im_i}\prod_{i=1}^r g_{l_i}(q; kl_i|s_i|) \neq \CI^{U(N)_k\times U(N)_{-k}}_H.
\end{equation}
The discrepancy comes from the $q^{-\sum_{i<j}|m_i-m_j| }$ factor in $\CI^{U(N)_k\times U(N)_{-k},\bfm}_H$ which counts the quantum corrected $\Delta-R_C$ charge of  monopole operators. This is an example that the Higgs branch got quantum corrections. This is different from ABJM theories.

If we look more carefully, equation \ref{eq:HiggsIndUNUNk} is actually the same as the Coulomb index of the 3d $U(N)$ SYM with $k$ fundamental hypermultiplets, so we conjecture that
\begin{equation}
\CM_H^{U(N)_k\times U(N)_{-k}} \simeq \CM_C^{U(N)~\mathrm{with}~k~\mathrm{hyper}},
\end{equation} 
up to some decoupled hypermultiplets when the theory is ugly.
Using results in \cite{Hanany:2011db, Bullimore:2015lsa}, we conjecture that the Higgs branch of $U(N)_k\times U(N)_{-k}$ CSM theory is the following intersection
\begin{equation}
\begin{split}
\CM_H^{U(N)_k\times U(N)_{-k}}& = S_{[k-N,N]}\cap\overline{\CO}_{[k]},\quad k\geq 2N,\\
\CM_H^{U(N)_k\times U(N)_{-k}}& = \bbC^2\times S_{[N,N-1]}\cap\overline{\CO}_{[2N-1]},\quad k= 2N-1,
\end{split}
\end{equation}
where $\overline{\CO}_{[k]}$ is the closure of the principal nilpotent orbit of $SU(k)$ and $S_{[k-N,N]}$ is the Slodowy slice corresponding to the nilpotent orbit $f=[k-N,N]$. Note that this does not mean that the $U(N)_k\times U(N)_{-k}$ CS theory is mirror to $U(N)$ SYM with $k$ fundamental-hypers as the Coulomb branch of the former is trivial while the Higgs branch of the later is non-trivial.

When $N_1\neq N_2$, the $\Delta-R_H$ charge of the monopole operator $V_{\bfm, \bfn}$ is
\begin{equation}
\frac{1}{2}\sum_{1\leq i\leq N_1}\sum_{1\leq j\leq N_2} |m_i-n_j|-\frac{1}{2}\sum_{1\leq i<j\leq N_1}|m_i-m_j|-\frac{1}{2}\sum_{1\leq i<j\leq N_2} |n_i-n_j|,
\end{equation}
which is again non-negative, so the Higgs index contribute only from monopoles with $0$ $\Delta-R_H$ charge dress with massless bosons. Without loss of generality one can assume $N_1<N_2$, and one finds that this requires  $\bfn$ to be  $\bfm$ with $N_2-N_1$ $0$'s added, then the computation of the Higgs index is the same as $N_1=N_2$.
In the end, the Higgs index when $N_1\neq N_2$ is
\begin{equation}
\CI^{U(N_1)_k\times U(N_2)_{-k}}_H=\prod_{i=1}^{\mathrm{min}(N_1,N_2)}\frac{1-q^{k-|N_1-N_2|+1-i}}{(1-q^i)(1-q^{\frac{k-|N_1-N_2|}{2}+1-i}w)(1-q^{\frac{k-|N_1-N_2|}{2}+1-i}w^{-1})},
\end{equation}
which is the same as the Higgs index of $U(N)_{\tilde{k}}\times U(N)_{-\tilde{k}}$ CSM theory with $N=\mathrm{min}(N_1,N_2)$ and the effective CS level $\tilde{k}=k-|N_1-N_2|$. This is because that at generic point of the moduli space, there are net $2|N_1-N_2|$ fermions gain negative masses, their one-loop contribution shift the CS level by $-|N_1-N_2|$ in total. And we also conjecture
\begin{equation}
\CM_H^{U(N_1)_k\times U(N_2)_{-k}}\simeq
\CM_C^{U(\mathrm{min}(N_1,N_2)))~\mathrm{with}~k-|N_1-N_2|~\mathrm{hyper}},
\end{equation}
up to some decoupled hypermultiplets when the theory is ugly.

One consistent check of our Higgs index is the duality between $U(N_1)_k\times U(N_2)_{-k}$ and $U(k-N_2)_k\times U(k-N_1)_{-k}$ theories. To make both side of the duality non-bad, $N_1$, $N_2$ and $k_1$ must satisfy
\begin{equation}
k-N_1-N_2 = 1,\ 0,\ \mathrm{or}\ -1.
\end{equation}
Without lose of generality, we can further assume $N_1<N_2$. When $k-N_1-N_2=-1$, we have
\begin{equation}
\CI^{U(N_1)_{N_1+N_2-1}\times U(N_2)_{-N_1-N_2+1}}_H=\frac{1}{ (1-q^{\frac{1}{2}}w )(1-q^{\frac{1}{2}}w^{-1} ) } \CI^{U(N_1-1)_{N_1+N_2-1}\times U(N_2-1)_{-N_1-N_2+1}}_H.
\end{equation}
On the other hand, when $k-N_1-N_2=1$, we have the opposite. The index confirms that $U(N_1)_k\times U(N_2)_{-k}$ and $U(k-N_2)_k\times U(k-N_1)_{-k}$ are dual theories up to a free hypermultiplet.

\subsubsection{The Coulomb index}
\label{eq:CBbifund}

From examples in section \ref{sec:exUN1UN2} one notices that the Coulomb indices are all $1$ up to the order of the expansion of $x$. Actually one can prove that the Coulomb indices are exactly $1$, and it is also a good example to illustrate that monopoles with negative $\Delta-R_C$ charges will not contribute to the Coulomb index after integration. 

Firstly consider the case $m_1>m_2>\cdots>m_{N_1}$ and $n_1>n_2>\cdots>n_{N_2}$,  the $\Delta-R_C$ charge of $V_{\bfm,\bfn}$ from equation \ref{eq:monopoleCharges} is 
\begin{equation}
-\frac{1}{2}\sum_{1\leq i<j\leq N_1}|m_i-m_j|-\frac{1}{2}\sum_{1\leq i<j \leq N_2} |n_i-n_j|,
\end{equation}
together with the $U(1)$ charges which are labelled by the index $\prod_i^{N_1} y^{km_i}_i \prod_{j=1}^{N_2} z_j^{-kn_j}$. To get a gauge invariant combination, one needs to compensate the index of the monopole with operators with gauge index  $\prod_i^{N_1} y^{-km_i}_i \prod_{j=1}^{N_2} z_j^{kn_j}$. The index \ref{eq:defN4hyp} indicates that each bosons or fermions in the bifundamental hyper with gauge charge either $x_iy_j^{-1}$ or $x^{-1}_iy_j$ have a $\Delta-R_C$ charge strictly greater than $|m_i-n_j|$.  Therefore the total $\Delta-R_C$ charge of the gauge singlet constructed from $V_{\bfm,\bfn}$ dressed with bosons and fermions is greater than
\begin{equation}
k\sum_{1\leq i,j \leq N} |m_i-n_j|-\sum_{1\leq i<j \leq N}(|m_i-m_j|+|n_i-n_j|),
\end{equation}
which is always positive when $k\geq1$. Therefore after taking $q\rightarrow 0$ limit, no gauge invariant operator will contribute to the Coulomb index.
One can also use similar argument to show that even when some $m_i$'s or $n_j$'s are equal, the monopole operator with dressing still does not contribute the the Coulomb index. This leaves only the contribution from vacuum only, hence the Coulomb index has to be $1$, indicating a trivial Coulomb branch which is expected as the free bifundamental hyper has no Coulomb branch in the beginning.

\subsection{$\CN=4$ CSM theories with linear quivers}
\label{sec:linearQuiverInd}

\begin{figure}

\begin{center}

\tikzset{every picture/.style={line width=0.75pt}} 

\begin{tikzpicture}[x=0.75pt,y=0.75pt,yscale=-1,xscale=1]

\draw   (20,50) .. controls (20,66.57) and (33.43,80) .. (50,80) .. controls (66.57,80) and (80,66.57) .. (80,50) .. controls (80,33.43) and (66.57,20) .. (50,20) .. controls (33.43,20) and (20,33.43) .. (20,50) -- cycle ;
\draw   (170,51) .. controls (170,34.43) and (156.57,21) .. (140,21) .. controls (123.43,21) and (110,34.43) .. (110,51) .. controls (110,67.57) and (123.43,81) .. (140,81) .. controls (156.57,81) and (170,67.57) .. (170,51) -- cycle ;
\draw    (81,51) -- (110,51) ;
\draw    (171,51) -- (200,51) ;
\draw   (509,50) .. controls (509,33.43) and (495.57,20) .. (479,20) .. controls (462.43,20) and (449,33.43) .. (449,50) .. controls (449,66.57) and (462.43,80) .. (479,80) .. controls (495.57,80) and (509,66.57) .. (509,50) -- cycle ;
\draw    (420,50) -- (449,50) ;
\draw [line width=1.5]  [dash pattern={on 1.69pt off 2.76pt}]  (390,50) -- (410,50) ;
\draw   (200,50) .. controls (200,66.57) and (213.43,80) .. (230,80) .. controls (246.57,80) and (260,66.57) .. (260,50) .. controls (260,33.43) and (246.57,20) .. (230,20) .. controls (213.43,20) and (200,33.43) .. (200,50) -- cycle ;
\draw   (350,50) .. controls (350,33.43) and (336.57,20) .. (320,20) .. controls (303.43,20) and (290,33.43) .. (290,50) .. controls (290,66.57) and (303.43,80) .. (320,80) .. controls (336.57,80) and (350,66.57) .. (350,50) -- cycle ;
\draw    (261,51) -- (290,51) ;
\draw    (351,50) -- (380,50) ;

\draw (25,42.4) node [anchor=north west][inner sep=0.75pt]    {$U( N_{1})_{k}$};
\draw (91,32.4) node [anchor=north west][inner sep=0.75pt]    {$h.$};
\draw (271,32.4) node [anchor=north west][inner sep=0.75pt]    {$h.$};
\draw (171,32.4) node [anchor=north west][inner sep=0.75pt]    {$t.h.$};
\draw (351,32.4) node [anchor=north west][inner sep=0.75pt]    {$t.h.$};
\draw (110,42.4) node [anchor=north west][inner sep=0.75pt]    {$U( N_{2})_{-k}$};
\draw (201,42.4) node [anchor=north west][inner sep=0.75pt]    {$U( N_{3})_{k}$};
\draw (288,42.4) node [anchor=north west][inner sep=0.75pt]    {$U( N_{4})_{-k}$};
\draw (459,42.4) node [anchor=north west][inner sep=0.75pt]    {$U( N_{l})_{( -)^{l+1} k}$};

\end{tikzpicture}

\tikzset{every picture/.style={line width=0.75pt}} 

\begin{tikzpicture}[x=0.75pt,y=0.75pt,yscale=-1,xscale=1]

\draw   (20,50) .. controls (20,66.57) and (33.43,80) .. (50,80) .. controls (66.57,80) and (80,66.57) .. (80,50) .. controls (80,33.43) and (66.57,20) .. (50,20) .. controls (33.43,20) and (20,33.43) .. (20,50) -- cycle ;
\draw   (170,51) .. controls (170,34.43) and (156.57,21) .. (140,21) .. controls (123.43,21) and (110,34.43) .. (110,51) .. controls (110,67.57) and (123.43,81) .. (140,81) .. controls (156.57,81) and (170,67.57) .. (170,51) -- cycle ;
\draw    (81,51) -- (110,51) ;
\draw    (171,51) -- (200,51) ;
\draw   (509,50) .. controls (509,33.43) and (495.57,20) .. (479,20) .. controls (462.43,20) and (449,33.43) .. (449,50) .. controls (449,66.57) and (462.43,80) .. (479,80) .. controls (495.57,80) and (509,66.57) .. (509,50) -- cycle ;
\draw    (420,50) -- (449,50) ;
\draw [line width=1.5]  [dash pattern={on 1.69pt off 2.76pt}]  (390,50) -- (410,50) ;
\draw   (200,50) .. controls (200,66.57) and (213.43,80) .. (230,80) .. controls (246.57,80) and (260,66.57) .. (260,50) .. controls (260,33.43) and (246.57,20) .. (230,20) .. controls (213.43,20) and (200,33.43) .. (200,50) -- cycle ;
\draw   (350,50) .. controls (350,33.43) and (336.57,20) .. (320,20) .. controls (303.43,20) and (290,33.43) .. (290,50) .. controls (290,66.57) and (303.43,80) .. (320,80) .. controls (336.57,80) and (350,66.57) .. (350,50) -- cycle ;
\draw    (261,51) -- (290,51) ;
\draw    (351,50) -- (380,50) ;

\draw (25,42.4) node [anchor=north west][inner sep=0.75pt]    {$U( N_{1})_{k}$};
\draw (81,32.4) node [anchor=north west][inner sep=0.75pt]    {$t.h.$};
\draw (261,32.4) node [anchor=north west][inner sep=0.75pt]    {$t.h.$};
\draw (181,32.4) node [anchor=north west][inner sep=0.75pt]    {$h.$};
\draw (355,32.4) node [anchor=north west][inner sep=0.75pt]    {$h.$};
\draw (110,42.4) node [anchor=north west][inner sep=0.75pt]    {$U( N_{2})_{-k}$};
\draw (201,42.4) node [anchor=north west][inner sep=0.75pt]    {$U( N_{3})_{k}$};
\draw (288,42.4) node [anchor=north west][inner sep=0.75pt]    {$U( N_{4})_{-k}$};
\draw (459,42.4) node [anchor=north west][inner sep=0.75pt]    {$U( N_{l})_{( -)^{l+1} k}$};

\end{tikzpicture}

\end{center}

\caption{\label{fig:N4linearQuiver} Upper: a linear quiver with $\CN=4$ CSM theories in the IR. Each node represents a $U(N_i)$ CS term with level $(-1)^{i+1} k$. Each line represents a bifundamental hyper or twisted hyper. Lower: its mirror quiver, with hypers and twisted hypers interchanged.}

\end{figure}
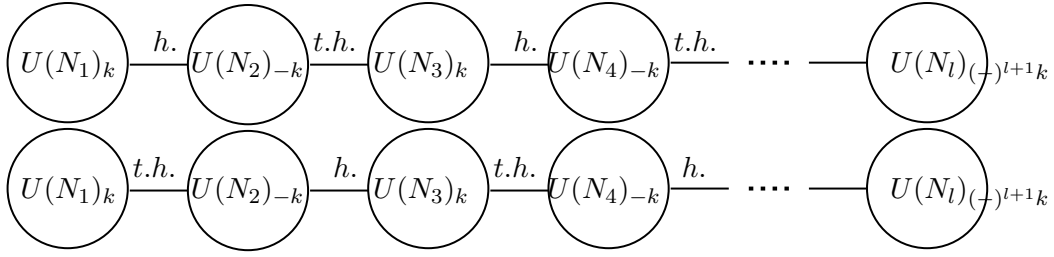

Now we term our attention to the index of CSM theories with linear quivers as shown in figure \ref{fig:N4linearQuiver} which is a generalization of Abelian theories discussed in section \ref{sec:U1U1U1} and \ref{sec:U1U1U1U1}. 
 For the theory to have $\CN=4$ SUSY in the IR, the CS levels should alternating between $k$ and $-k$, while the matters should alternating between bifundamental hypermultiplets and twisted hypermultiplets. Apparently this theory is mirror to the theory with all hypers and twisted hypers  interchanged as shown in the lower part of figure \ref{fig:N4linearQuiver}. Moreover,  the theory is self-mirror when $l$ is an odd number and $N_{i}=N_{l+1-i}$, and this provides a systematical way of constructing self-mirror theories.

The index of the upper theory in figure \ref{fig:N4linearQuiver} is
\begin{equation}
\begin{split}
\label{eq:linearfull}
\CI 
=&\sum_{\bfm^{(1)},\cdots,\bfm^{(l)}}\prod_{i=1}^l\frac{1}{W(\bfm^{(l)}) }\int \prod_{i=1}^l \frac{d\bfy^{(i)}}{2\pi i\bfy^{(i)}} 
\prod_{i=1}^l \prod_{j=1}^{N_i} \left( u^{(i)}(y^{(i)}_j)^{(-1)^{i+1} k }\right)^{ m^{(i)}_j}
\\
&\times Z_{vec}(\{\bfy^{(1)},\bfm^{(1)}\}) 
\prod_{i=2}^{l} Z_{hyp}(\{\bfy^{(i-1)},\bfm^{(i-1)}\},\{\bfy^{(i)},\bfm^{(i)} \}  ; t^{(-1)^{i}} ) Z_{vec}(\{\bfy^{(i)},\bfm^{(i)} \}),
\end{split}
\end{equation}
with $\bfy^{(i)}$, $\bfm^{(i)}$ and $u^{(i)}$ being the collection of the gauge fugacities, monopole charges and fugacity of the topological $U(1)$ corresponding to the $U(N_i)$ nodes. Usually it is computationally heavy to work out the full index order by order, however, in the next section we will present a simple gluing formula for the Higgs and Coulomb index.

\subsubsection{The gluing formula of Higgs and Coulomb indices}

As shown in the theory with only $1$ bi-fundamental hypermultiplet, if one can figure out monopole operators contributes to Higgs/Coulomb indices before performing the integration in the index, it will greatly simplify the computation and final results. A general rule of thumb is that a hypermultiplet contributes positively to $\Delta-R_H$ and $0$ to $\Delta-R_C$, a twisted hypermulitplet contributes $0$ to $\Delta-R_C$ and positively to $\Delta-R_H$, and finally an $\CN=2$ vector multiplet both negatively to $\Delta-R_C$ and $\Delta-R_H$ for the same amount. If the total quantum $\Delta-R_H$ ($\Delta-R_C$ ) charge of any monopole is always non-negative, one can then take the $\tilde{q}\rightarrow 0$ ($q\rightarrow 0$) limit in advance and the integrand of the index is greatly simplified. If the total quantum $\Delta-R_H$ ($\Delta-R_C$) charge of certain monopole is negative, one can always exclude the contribution of such monopole in the full index using the argument in \ref{eq:CBbifund}.

Now we compute the Higgs index of the upper theory in figure \ref{fig:N4linearQuiver}. When $l=2r$ is a even number, the theory can be viewed as $r$ $U(N_i)\times U(N_{i+1})$ CSM theories with $1$ bifundamental hypermultiplet glued by $r-1$ twisted bifundamental hypermultiplets, and the quantum corrected $\Delta-R_H$ of a monopole is roughly $r$ copies of the charge in \ref{eq:DeltaRHcharge}, so is always non negative. Therefore the monopole $V_{\bfm^1,\bfm^2,\cdots,\bfm^{2r-1},\bfm^{2r}}$ may contribute to the Higgs index if its monopole charges satisfying the following constraints
\begin{equation}
\bfm^{2k-1}=\bfm^{2k},\quad k=1,2,\cdots,r.
\end{equation}
 Here by $\bfm^{2k-1}=\bfm^{2k}$ we mean that $\bfm^{2k-1}=\bfm^{2k}$ when $N_{2k-1}=N_{2k}$, while $\bfm^{2k-1}$ is the same as $\bfm^{2k}$ with $N_1-N_2$ $0$'s inserted in the correct position when $N_{2k-1}>N_{2k}$ and vice versa. When $l=2r+1$ being an odd number, there is an extra contribution to $\Delta-R_H$ from the $\CN=2$ vector multiplet of $U(N_{2r+1})$, we see that gauge invariance further fix $\bfm^{2r+1}$ to be $\bf 0$.

Therefore, the total Higgs index looks like the Higgs indices of $U(N_i)_k\times U(N_{i+1})_{-k}$ theories glued by the twisted hypermultiplets. Denoting the Higgs index of the $U(N_i)_k\times U(N_{i+1})_{-k}$ theory with background monopole charge $(\bfm,\bfm)$ by $\CI^{N_i,N_{i+1},k,\bfm }_H$ which is defined in equation \ref{eq:bifundkm}. The Higgs index of the upper theory of figure \ref{fig:N4linearQuiver} is then computed by the following gluing formula for odd or even number of gauge groups,
\begin{equation}
\label{eq:Higgslinear}
\begin{split}
\CI_H^{U(N_1)_k\times\cdots U(N_{2r+1})_{k}}  
&=\sum_{\bfm^{1},\bfm^{3}, \cdots,\bfm^{2r-1} } 
\CI_H^{N_1,N_2,k,\bfm^{1}} q^{\frac{ |m^{1}_{i_1 } - m^{ 3 }_{i_3 } |}{2} } \CI_H^{N_3,N_4,k,\bfm^{3}} \times \cdots \times  \CI_H^{N_{2r-1},N_{2r},k,\bfm^{2r-1}} q^{\frac{|m_{i_{2r-1}}^{2r-1}|}{2} }, \\
  \CI_H^{U(N_1)_k\times\cdots U(N_{2r})_{k}} 
&=\sum_{\bfm^{1},\bfm^3 \cdots,\bfm^{2r-1} } 
\CI_H^{N_1,N_2,k,\bfm^{1}} q^{\frac{ |m^{1}_{i_1} - m^{3 }_{i_3} |}{2} } \CI_H^{N_3,N_4,k,\bfm^{3}} \times \cdots \times  \CI_H^{N_{2r-1},N_{2r},k,\bfm^{2r-1}}.
\end{split}
\end{equation}
Here $\sum_{\bfm^i}$ is the collective notation for $\sum_{m^i_1\geq m^i_2\geq\cdots\geq m^i_{N_i}}$ which is the sum over all monopoles of $U(N_i)$.
The term $q^{\frac{ |m^{(2s-1)}_i - m^{(2s+1) }_j |}{2} }$ is the quantum correction to the $\Delta-R_H$ charge from the $s$-th twisted hypermultiplet, and because of this term, the total Higgs index is not simply a product of the Higgs index of all hypers. 

Similarly the Coulomb branch can also be computed by gluing among the Coulomb branch from twisted hypers between $U(2i)_{-k}\times U(2i+1)_k$,
\begin{equation}
\label{eq:Coulomblinear}
\begin{split}
 \CI_C^{U(N_1)_k\times\cdots U(N_{2r+1})_{k}} 
= & \sum_{\bfm^{2}, \cdots,\bfm^{2r} } 
\tilde{q}^{ \frac{|m^{2}_{i_2}| }{2} } \CI_C^{N_2,N_3,k,\bfm^{2}} \tilde{q}^{\frac{ |m^{2}_{i_2} - m^{4 }_{i_4} |}{2} } \CI_C^{N_4,N_5,k,\bfm^{4}} \times \cdots \times  \CI_C^{N_{2r},N_{2r+1},k,\bfm^{2r} },  \\
 \CI_C^{U(N_1)_k\times\cdots U(N_{2r+2})_{k}} 
=&\sum_{\bfm^{2}, \cdots,\bfm^{2r} } 
\tilde{q}^{ \frac{|m^{2}_{i_2}| }{2} } \CI_C^{N_2,N_3,k,\bfm^{2}} \tilde{q}^{\frac{ |m^{2}_{i_2} - m^{4 }_{i_4} |}{2} } \CI_C^{N_4,N_5,k,\bfm^{4}} \times \cdots \times  \CI_C^{N_{2r},N_{2r+1},k,\bfm^{2r} \tilde{q}^{ \frac{|m^{2r}_{i_{2r}}| }{2} }}.
\end{split}
\end{equation}

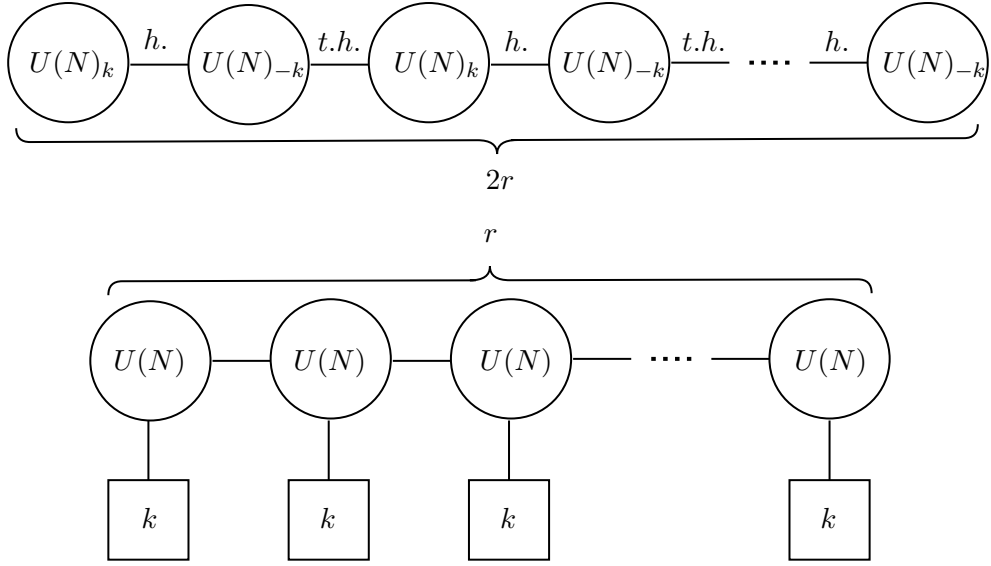
\begin{figure}
\begin{center}

\tikzset{every picture/.style={line width=0.75pt}} 

\begin{tikzpicture}[x=0.75pt,y=0.75pt,yscale=-1,xscale=1]

\draw   (20,50) .. controls (20,66.57) and (33.43,80) .. (50,80) .. controls (66.57,80) and (80,66.57) .. (80,50) .. controls (80,33.43) and (66.57,20) .. (50,20) .. controls (33.43,20) and (20,33.43) .. (20,50) -- cycle ;
\draw   (170,51) .. controls (170,34.43) and (156.57,21) .. (140,21) .. controls (123.43,21) and (110,34.43) .. (110,51) .. controls (110,67.57) and (123.43,81) .. (140,81) .. controls (156.57,81) and (170,67.57) .. (170,51) -- cycle ;
\draw    (81,51) -- (110,51) ;
\draw    (171,51) -- (200,51) ;
\draw   (509,50) .. controls (509,33.43) and (495.57,20) .. (479,20) .. controls (462.43,20) and (449,33.43) .. (449,50) .. controls (449,66.57) and (462.43,80) .. (479,80) .. controls (495.57,80) and (509,66.57) .. (509,50) -- cycle ;
\draw    (420,50) -- (449,50) ;
\draw [line width=1.5]  [dash pattern={on 1.69pt off 2.76pt}]  (390,50) -- (410,50) ;
\draw   (200,50) .. controls (200,66.57) and (213.43,80) .. (230,80) .. controls (246.57,80) and (260,66.57) .. (260,50) .. controls (260,33.43) and (246.57,20) .. (230,20) .. controls (213.43,20) and (200,33.43) .. (200,50) -- cycle ;
\draw   (350,50) .. controls (350,33.43) and (336.57,20) .. (320,20) .. controls (303.43,20) and (290,33.43) .. (290,50) .. controls (290,66.57) and (303.43,80) .. (320,80) .. controls (336.57,80) and (350,66.57) .. (350,50) -- cycle ;
\draw    (261,51) -- (290,51) ;
\draw    (351,50) -- (380,50) ;
\draw   (121,200) .. controls (121,183.43) and (107.57,170) .. (91,170) .. controls (74.43,170) and (61,183.43) .. (61,200) .. controls (61,216.57) and (74.43,230) .. (91,230) .. controls (107.57,230) and (121,216.57) .. (121,200) -- cycle ;
\draw    (122,200) -- (151,200) ;
\draw   (460,199) .. controls (460,182.43) and (446.57,169) .. (430,169) .. controls (413.43,169) and (400,182.43) .. (400,199) .. controls (400,215.57) and (413.43,229) .. (430,229) .. controls (446.57,229) and (460,215.57) .. (460,199) -- cycle ;
\draw    (371,199) -- (400,199) ;
\draw [line width=1.5]  [dash pattern={on 1.69pt off 2.76pt}]  (341,199) -- (361,199) ;
\draw   (151,199) .. controls (151,215.57) and (164.43,229) .. (181,229) .. controls (197.57,229) and (211,215.57) .. (211,199) .. controls (211,182.43) and (197.57,169) .. (181,169) .. controls (164.43,169) and (151,182.43) .. (151,199) -- cycle ;
\draw   (301,199) .. controls (301,182.43) and (287.57,169) .. (271,169) .. controls (254.43,169) and (241,182.43) .. (241,199) .. controls (241,215.57) and (254.43,229) .. (271,229) .. controls (287.57,229) and (301,215.57) .. (301,199) -- cycle ;
\draw    (212,200) -- (241,200) ;
\draw    (302,199) -- (331,199) ;
\draw    (90,230) -- (90,260) ;
\draw   (70,260) -- (110,260) -- (110,300) -- (70,300) -- cycle ;
\draw    (430,230) -- (430,260) ;
\draw   (410,260) -- (450,260) -- (450,300) -- (410,300) -- cycle ;
\draw    (180,230) -- (180,260) ;
\draw   (160,260) -- (200,260) -- (200,300) -- (160,300) -- cycle ;
\draw    (270,230) -- (270,260) ;
\draw   (250,260) -- (290,260) -- (290,300) -- (250,300) -- cycle ;
\draw   (24,82.97) .. controls (24.02,87.64) and (26.36,89.96) .. (31.03,89.94) -- (254.03,89.01) .. controls (260.7,88.98) and (264.04,91.29) .. (264.06,95.96) .. controls (264.04,91.29) and (267.36,88.95) .. (274.03,88.92)(271.03,88.93) -- (497.03,87.99) .. controls (501.7,87.97) and (504.02,85.63) .. (504,80.96) ;
\draw   (451,165.97) .. controls (450.99,161.3) and (448.65,158.98) .. (443.98,158.99) -- (270.48,159.44) .. controls (263.81,159.46) and (260.47,157.14) .. (260.46,152.47) .. controls (260.47,157.14) and (257.15,159.48) .. (250.48,159.49)(253.48,159.49) -- (76.98,159.95) .. controls (72.31,159.96) and (69.99,162.3) .. (70,166.97) ;

\draw (29,42.4) node [anchor=north west][inner sep=0.75pt]    {$U( N)_{k}$};
\draw (86,31.4) node [anchor=north west][inner sep=0.75pt]    {$h.$};
\draw (266,32.4) node [anchor=north west][inner sep=0.75pt]    {$h.$};
\draw (173,32.4) node [anchor=north west][inner sep=0.75pt]    {$t.h.$};
\draw (355,32.4) node [anchor=north west][inner sep=0.75pt]    {$t.h.$};
\draw (115,42.4) node [anchor=north west][inner sep=0.75pt]    {$U( N)_{-k}$};
\draw (426,32.4) node [anchor=north west][inner sep=0.75pt]    {$h.$};
\draw (71,192.4) node [anchor=north west][inner sep=0.75pt]    {$U( N)$};
\draw (161,192.4) node [anchor=north west][inner sep=0.75pt]    {$U( N)$};
\draw (254,192.4) node [anchor=north west][inner sep=0.75pt]    {$U( N)$};
\draw (411,192.4) node [anchor=north west][inner sep=0.75pt]    {$U( N)$};
\draw (85,272.4) node [anchor=north west][inner sep=0.75pt]    {$k$};
\draw (424,272.4) node [anchor=north west][inner sep=0.75pt]    {$k$};
\draw (174,272.4) node [anchor=north west][inner sep=0.75pt]    {$k$};
\draw (264,272.4) node [anchor=north west][inner sep=0.75pt]    {$k$};
\draw (211,42.4) node [anchor=north west][inner sep=0.75pt]    {$U( N)_{k}$};
\draw (295,42.4) node [anchor=north west][inner sep=0.75pt]    {$U( N)_{-k}$};
\draw (455,42.4) node [anchor=north west][inner sep=0.75pt]    {$U( N)_{-k}$};
\draw (257,102.4) node [anchor=north west][inner sep=0.75pt]    {$2r$};
\draw (256,132.4) node [anchor=north west][inner sep=0.75pt]    {$r$};

\end{tikzpicture}

\end{center}
\caption{\label{fig:CBCS=YM}The index implies that the Higgs branch of the upper CSM theory is the same as the Coulomb index of the lower SYM theory.}
\end{figure}

Actually, using the fact that $\CI^{N_i,N_{i+1},k,\bfm }_H$ is the same as the Coulomb index of $\CN=4$ $U(\mathrm{min}(N_i, N_{i+1}))$ SYM with $k-|N_1,N_2|$ fundamental hypermultiplets with monopole charge $\bfm$, the gluing formula \ref{eq:Higgslinear} implies that the Higgs branch is the same as the Coulomb branch of a $\CN=4$ SYM theory with gauge group $U(\mathrm{min}(N_1, N_{2})\times U(\mathrm{min}(N_3,N_4))\times\cdots\times U(\mathrm{min}(N_{2r-1}, N_{2r})$ with bifundamental hypermultiplets between adjacent gauge groups, moreover each gauge group is dressed with $k-|N_{2k-1}- N_{2k}|$ fundamental hypermutliplets (see figure \ref{fig:CBCS=YM} for an example). Similar statement can also be made for the Coulomb branch.

{\bf Example:} as an example of our gluing formula \ref{eq:Higgslinear} and \ref{eq:Coulomblinear}, we compute the Higgs of the $U(1)_1\times U(1)_{-1}\times U(1)_1\times U(1)_{-1}$ theory discussed before. According to equation \ref{eq:bifundkm}, the Higgs index of the $U(1)_1\times U(1)_{-1}$ theory at monopole charge $(m,m)$ is
\begin{equation}
\CI^{1,1,1,m}_H = w^m g_1(q; |m|) = \frac{ w^m q^{|m|/2} }{ 1-q }.
\end{equation}
Then gluing formula \ref{eq:Higgslinear} tells us that the Higgs index of the full $U(1)_1\times U(1)_{-1}\times U(1)_1\times U(1)_{-1}$ theory is
\begin{equation}
\begin{split}
\CI_H & = \sum_{m,n\in\bbZ} \frac{ w_1^m q^{|m|/2} }{ 1-q } q^{|m-n|/2 } \frac{ w_2^n q^{|n|/2} }{ 1-q } \\
& = \frac{ 1 + 2q - q^2 \left( w_1+w_1^{-1} +w_2 +w_2^{-1} + w_1w_2 + (w_1w_2)^{-1} \right) +2 q^3 +q^4 } { (1-q w_1 ) (1-q w_1^{-1} ) (1 - q w_2 )  (1 - q w_2^{-1} ) ( 1 - q w_1w_2 )  ( 1 - q (w_1w_2)^{-1} )}.
\end{split}
\end{equation}
Direct computation or using the identity in appendix \ref{app:sec:mirrorId} one can see that this is the same as the Higgs index of $U(1)$ SYM with $3$ flavors.

\subsubsection{Duality between $\CN=4$ CSM theories and usual $\mathcal{N}=4$ gauge theory}
\label{sec:mirrorLinearQuiver}

One application of the gluing formula \ref{eq:Higgslinear} and \ref{eq:Coulomblinear} and the full superconformal index is to predict and check 3d mirror of $\CN=4$ CSM theories. Here we discuss  mirrors of the CSM linear quiver theories which generalize the results in \cite{Jafferis:2008em,Hosomichi:2008jd}.

\begin{figure}
\begin{center}

\tikzset{every picture/.style={line width=0.75pt}} 

\begin{tikzpicture}[x=0.75pt,y=0.75pt,yscale=-1,xscale=1]

\draw   (20,50) .. controls (20,66.57) and (33.43,80) .. (50,80) .. controls (66.57,80) and (80,66.57) .. (80,50) .. controls (80,33.43) and (66.57,20) .. (50,20) .. controls (33.43,20) and (20,33.43) .. (20,50) -- cycle ;
\draw   (170,51) .. controls (170,34.43) and (156.57,21) .. (140,21) .. controls (123.43,21) and (110,34.43) .. (110,51) .. controls (110,67.57) and (123.43,81) .. (140,81) .. controls (156.57,81) and (170,67.57) .. (170,51) -- cycle ;
\draw    (81,51) -- (110,51) ;
\draw    (171,51) -- (200,51) ;
\draw   (509,50) .. controls (509,33.43) and (495.57,20) .. (479,20) .. controls (462.43,20) and (449,33.43) .. (449,50) .. controls (449,66.57) and (462.43,80) .. (479,80) .. controls (495.57,80) and (509,66.57) .. (509,50) -- cycle ;
\draw    (420,50) -- (449,50) ;
\draw [line width=1.5]  [dash pattern={on 1.69pt off 2.76pt}]  (390,50) -- (410,50) ;
\draw   (200,50) .. controls (200,66.57) and (213.43,80) .. (230,80) .. controls (246.57,80) and (260,66.57) .. (260,50) .. controls (260,33.43) and (246.57,20) .. (230,20) .. controls (213.43,20) and (200,33.43) .. (200,50) -- cycle ;
\draw   (350,50) .. controls (350,33.43) and (336.57,20) .. (320,20) .. controls (303.43,20) and (290,33.43) .. (290,50) .. controls (290,66.57) and (303.43,80) .. (320,80) .. controls (336.57,80) and (350,66.57) .. (350,50) -- cycle ;
\draw    (261,51) -- (290,51) ;
\draw    (351,50) -- (380,50) ;
\draw   (131,140) .. controls (131,123.43) and (117.57,110) .. (101,110) .. controls (84.43,110) and (71,123.43) .. (71,140) .. controls (71,156.57) and (84.43,170) .. (101,170) .. controls (117.57,170) and (131,156.57) .. (131,140) -- cycle ;
\draw    (132,140) -- (161,140) ;
\draw   (470,139) .. controls (470,122.43) and (456.57,109) .. (440,109) .. controls (423.43,109) and (410,122.43) .. (410,139) .. controls (410,155.57) and (423.43,169) .. (440,169) .. controls (456.57,169) and (470,155.57) .. (470,139) -- cycle ;
\draw    (381,139) -- (410,139) ;
\draw [line width=1.5]  [dash pattern={on 1.69pt off 2.76pt}]  (351,139) -- (371,139) ;
\draw   (161,139) .. controls (161,155.57) and (174.43,169) .. (191,169) .. controls (207.57,169) and (221,155.57) .. (221,139) .. controls (221,122.43) and (207.57,109) .. (191,109) .. controls (174.43,109) and (161,122.43) .. (161,139) -- cycle ;
\draw   (311,139) .. controls (311,122.43) and (297.57,109) .. (281,109) .. controls (264.43,109) and (251,122.43) .. (251,139) .. controls (251,155.57) and (264.43,169) .. (281,169) .. controls (297.57,169) and (311,155.57) .. (311,139) -- cycle ;
\draw    (222,140) -- (251,140) ;
\draw    (312,139) -- (341,139) ;
\draw    (100,170) -- (100,200) ;
\draw   (80,200) -- (120,200) -- (120,240) -- (80,240) -- cycle ;
\draw    (440,170) -- (440,200) ;
\draw   (420,200) -- (460,200) -- (460,240) -- (420,240) -- cycle ;
\draw    (190,170) -- (190,200) ;
\draw   (170,200) -- (210,200) -- (210,240) -- (170,240) -- cycle ;
\draw    (280,170) -- (280,200) ;
\draw   (260,200) -- (300,200) -- (300,240) -- (260,240) -- cycle ;

\draw (29,42.4) node [anchor=north west][inner sep=0.75pt]    {$U( 1)_{1}$};
\draw (81,32.4) node [anchor=north west][inner sep=0.75pt]    {$t.h.$};
\draw (261,32.4) node [anchor=north west][inner sep=0.75pt]    {$t.h.$};
\draw (181,32.4) node [anchor=north west][inner sep=0.75pt]    {$h.$};
\draw (355,32.4) node [anchor=north west][inner sep=0.75pt]    {$h.$};
\draw (119,42.4) node [anchor=north west][inner sep=0.75pt]    {$U( 1)_{-1}$};
\draw (211,40.4) node [anchor=north west][inner sep=0.75pt]    {$U( 1)_{1}$};
\draw (299,42.4) node [anchor=north west][inner sep=0.75pt]    {$U( 1)_{-1}$};
\draw (456,42.4) node [anchor=north west][inner sep=0.75pt]    {$U( 1)_{-1}$};
\draw (420,32.4) node [anchor=north west][inner sep=0.75pt]    {$t.h.$};
\draw (81,132.4) node [anchor=north west][inner sep=0.75pt]    {$U( 1)$};
\draw (174,132.4) node [anchor=north west][inner sep=0.75pt]    {$U( 1)$};
\draw (264,132.4) node [anchor=north west][inner sep=0.75pt]    {$U( 1)$};
\draw (421,132.4) node [anchor=north west][inner sep=0.75pt]    {$U( 1)$};
\draw (95,212.4) node [anchor=north west][inner sep=0.75pt]    {$2$};
\draw (434,212.4) node [anchor=north west][inner sep=0.75pt]    {$2$};
\draw (184,212.4) node [anchor=north west][inner sep=0.75pt]    {$1$};
\draw (274,212.4) node [anchor=north west][inner sep=0.75pt]    {$1$};

\end{tikzpicture}

\end{center}
\caption{\label{fig:U12r} Upper: $U(1)^{2r}$ CSM theories with alternating twisted hypers and hypers. Lower: the mirror theory is an $\CN=4$ YM theory with linear quiver. All nodes except the first and last are dressed with one extra hyper. The first and last nodes are dressed with two hypers. }
\end{figure}
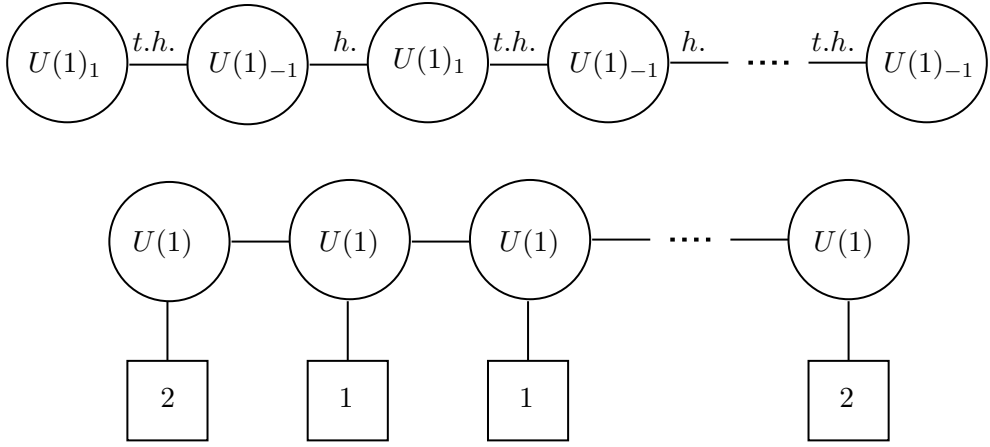
Starting with the CSM theories shown in the upper part of figure \ref{fig:U12r} with gauge group $U(1)^{2r}$ and level $\pm1$. Following the glueing formula, its Higgs branch is
\begin{equation}
\CI_H = \sum_{m_1,\cdots,m_{r-1}}q^{|m_1|/2}\frac{w_1^{m_1} q^{|m_1|/2}} {1-q} q^{|m_1-m_2|/2} \frac{w_2^{m_2} q^{|m_2|/2}} {1-q}\cdots q^{|m_{r-2}-m_{r-1}|/2} \frac{w_{r-1}^{m_{r-1}} q^{|m_{r-1}|/2}}{1-q}q^{|m_{r-1}|/2},
\end{equation}
which is exactly the same as the Coulomb branch of $\CN=4$ $U(1)^{r-1}$ YM theory dressed with flavors as depicted in the lower part of figure \ref{fig:U12r}.

The matching between the Coulomb branch of the CSM theory and the Higgs branch of its mirror is less trivial. According to the gluing formula, the Coulomb index of the CSM theory is
\begin{equation}
\CI_C = \sum_{m_1,\cdots,m_{r}}\frac{ \tq^{|m_1|/2}} {1-\tq} \tq^{|m_1-m_2|/2} \frac{  \tq^{|m_2|/2}} {1-\tq}\cdots \tq^{|m_{r-1}-m_{r}|/2} \frac{  \tq^{|m_{r}|/2}}{1-\tq},
\end{equation}
where we set the fugacities of topological symmetry to $1$ for simplicity.
On the other hand, the Higgs index of its mirror is
\begin{equation}
\CI_H^{mirror}=\int \prod_{i=1}^{r-1}\frac{dz_i}{2\pi i z_i}(1-\tq)^{r-1}\frac{1}{(1-\tq^{\frac{1}{2}} z_1^\pm)^2} \frac{1}{(1-\tq^{\frac{1}{2}} (z_2/z_1)^\pm) }\frac{1}{(1-\tq^{\frac{1}{2}} z_2^\pm)} \frac{1}{(1-\tq^{\frac{1}{2}} (z_3/z_2)^\pm) }\cdots \frac{1}{(1-\tq^{\frac{1}{2}} z_{r-1}^\pm)^2}.
\end{equation}
One can show that the unrefined index $\CI_C$ and $\CI_H^{mirror}$ are indeed equal
\begin{equation}
\CI_C = \CI_H^{mirror}.
\end{equation}
The details of this equality are summarized in appendix \ref{app:sec:mirrorId}.

One can also check that the full indices of the CSM theory and its mirror are related by $t\rightarrow t^{-1}$. For example, using \ref{eq:linearfull} the unrefined $\CN=4$ index of the $U(1)^6$ CSM theory is
\begin{equation}
\begin{split}
\CI^{U(1)^6} = &
1+\left(2t^2+\frac{7}{t^2}\right) x + \left(4t^3+\frac{8}{t^3}\right) x^{3/2} + \left(5t^4 - 4 +\frac{26}{t^4}\right) x^2 \\
& + \left( 8 t^5 +8 t - \frac{8}{t} +\frac{40}{t^5} \right) x^{5/2} + \left(14t^6 - 6t^2 - \frac{42}{t^2} + \frac{ 88 }{t^6} \right) x^3 +\cdots. 
\end{split}
\end{equation}
The $\CN=4$ index of the mirror YM theory is
\begin{equation}
\begin{split}
\CI^{mirror} = &
1+\left( 2t^{-2}+7t^2 \right) x + \left(4t^{-3}+{8 t^3}\right) x^{3/2} + \left(5t^{-4} - 4 + {26 t^4}\right) x^2 \\
& + \left( 8 t^{-5} +8 t^{-1}  -  8 t  + {40 t^5} \right) x^{5/2} + \left(14t^{-6} - 6t^{-2} -  {42 t^2} +   88  t^6 \right) x^3 +\cdots. 
\end{split}
\end{equation}
Indeed we have
\begin{equation}
\CI^{U(1)^6}\xrightarrow{t\rightarrow t^{-1}} \CI^{mirror},
\end{equation}
which gives strong evidence for the mirror symmetry.

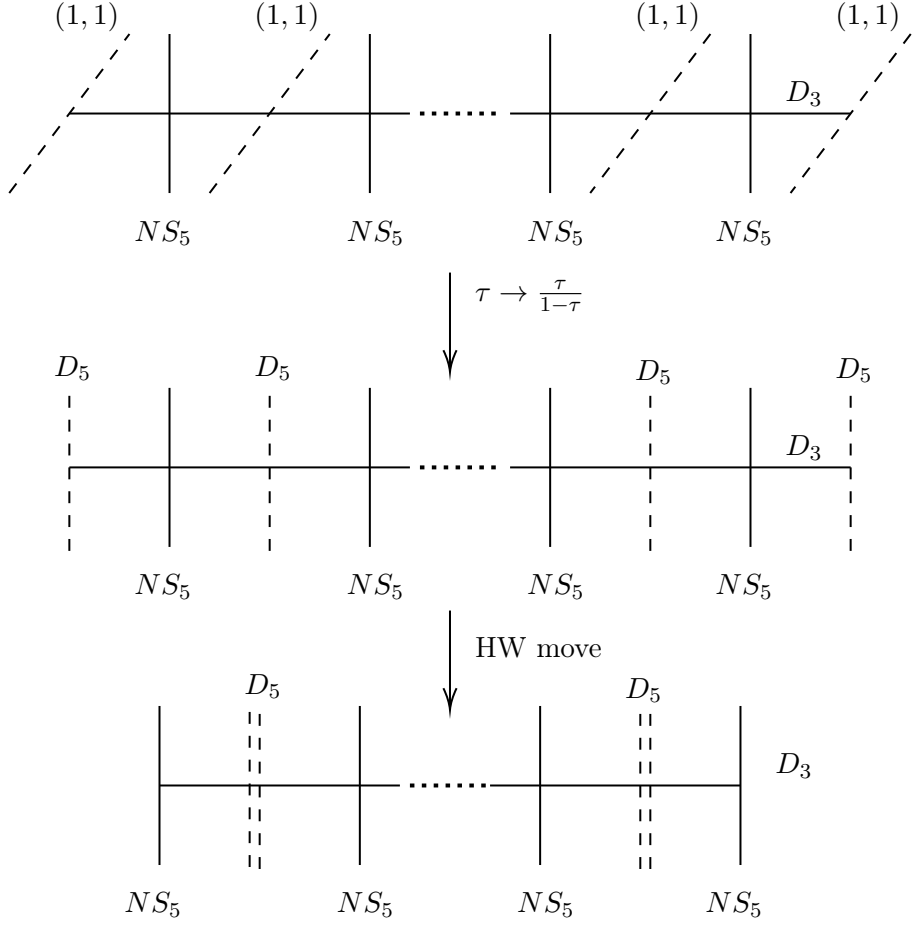
\begin{figure}
\begin{center}

\tikzset{every picture/.style={line width=0.75pt}} 

\begin{tikzpicture}[x=0.75pt,y=0.75pt,yscale=-1,xscale=1]

\draw    (50,80) -- (220,80) ;
\draw  [dash pattern={on 4.5pt off 4.5pt}]  (20,120) -- (80,40) ;
\draw    (100,40) -- (100,120) ;
\draw  [dash pattern={on 4.5pt off 4.5pt}]  (120,120) -- (180,40) ;
\draw    (200,40) -- (200,120) ;
\draw [line width=1.5]  [dash pattern={on 1.69pt off 2.76pt}]  (225,80) -- (265,80) ;
\draw    (440,80) -- (270,80) ;
\draw  [dash pattern={on 4.5pt off 4.5pt}]  (470,40) -- (410,120) ;
\draw    (390,120) -- (390,40) ;
\draw  [dash pattern={on 4.5pt off 4.5pt}]  (370,40) -- (310,120) ;
\draw    (290,120) -- (290,40) ;
\draw    (240,160) -- (240,208) ;
\draw [shift={(240,210)}, rotate = 270] [color={rgb, 255:red, 0; green, 0; blue, 0 }  ][line width=0.75]    (10.93,-3.29) .. controls (6.95,-1.4) and (3.31,-0.3) .. (0,0) .. controls (3.31,0.3) and (6.95,1.4) .. (10.93,3.29)   ;
\draw    (50,258) -- (220,258) ;
\draw  [dash pattern={on 4.5pt off 4.5pt}]  (50,300) -- (50,220) ;
\draw    (100,218) -- (100,298) ;
\draw    (200,218) -- (200,298) ;
\draw [line width=1.5]  [dash pattern={on 1.69pt off 2.76pt}]  (225,258) -- (265,258) ;
\draw    (440,258) -- (270,258) ;
\draw    (390,298) -- (390,218) ;
\draw    (290,298) -- (290,218) ;
\draw  [dash pattern={on 4.5pt off 4.5pt}]  (150,300) -- (150,220) ;
\draw  [dash pattern={on 4.5pt off 4.5pt}]  (340,300) -- (340,220) ;
\draw  [dash pattern={on 4.5pt off 4.5pt}]  (440,300) -- (440,220) ;
\draw    (240,330) -- (240,378) ;
\draw [shift={(240,380)}, rotate = 270] [color={rgb, 255:red, 0; green, 0; blue, 0 }  ][line width=0.75]    (10.93,-3.29) .. controls (6.95,-1.4) and (3.31,-0.3) .. (0,0) .. controls (3.31,0.3) and (6.95,1.4) .. (10.93,3.29)   ;
\draw    (95,418) -- (215,418) ;
\draw  [dash pattern={on 4.5pt off 4.5pt}]  (140,459) -- (140,379) ;
\draw    (95,378) -- (95,458) ;
\draw    (195,378) -- (195,458) ;
\draw [line width=1.5]  [dash pattern={on 1.69pt off 2.76pt}]  (220,418) -- (260,418) ;
\draw    (385,418) -- (260,418) ;
\draw    (385,458) -- (385,378) ;
\draw    (285,458) -- (285,378) ;
\draw  [dash pattern={on 4.5pt off 4.5pt}]  (145,460) -- (145,380) ;
\draw  [dash pattern={on 4.5pt off 4.5pt}]  (335,460) -- (335,380) ;
\draw  [dash pattern={on 4.5pt off 4.5pt}]  (340,460) -- (340,380) ;

\draw (41,22.4) node [anchor=north west][inner sep=0.75pt]    {$( 1,1)$};
\draw (141,22.4) node [anchor=north west][inner sep=0.75pt]    {$( 1,1)$};
\draw (331,22.4) node [anchor=north west][inner sep=0.75pt]    {$( 1,1)$};
\draw (431,22.4) node [anchor=north west][inner sep=0.75pt]    {$( 1,1)$};
\draw (81,132.4) node [anchor=north west][inner sep=0.75pt]    {$NS_{5}$};
\draw (187,132.4) node [anchor=north west][inner sep=0.75pt]    {$NS_{5}$};
\draw (277,132.4) node [anchor=north west][inner sep=0.75pt]    {$NS_{5}$};
\draw (371,132.4) node [anchor=north west][inner sep=0.75pt]    {$NS_{5}$};
\draw (406,62.4) node [anchor=north west][inner sep=0.75pt]    {$D_{3}$};
\draw (251,162.4) node [anchor=north west][inner sep=0.75pt]    {$\tau \rightarrow \frac{\tau }{1-\tau }$};
\draw (41,200.4) node [anchor=north west][inner sep=0.75pt]    {$D_{5}$};
\draw (81,310.4) node [anchor=north west][inner sep=0.75pt]    {$NS_{5}$};
\draw (187,310.4) node [anchor=north west][inner sep=0.75pt]    {$NS_{5}$};
\draw (277,310.4) node [anchor=north west][inner sep=0.75pt]    {$NS_{5}$};
\draw (371,310.4) node [anchor=north west][inner sep=0.75pt]    {$NS_{5}$};
\draw (406,240.4) node [anchor=north west][inner sep=0.75pt]    {$D_{3}$};
\draw (141,200.4) node [anchor=north west][inner sep=0.75pt]    {$D_{5}$};
\draw (331,202.4) node [anchor=north west][inner sep=0.75pt]    {$D_{5}$};
\draw (431,201.4) node [anchor=north west][inner sep=0.75pt]    {$D_{5}$};
\draw (251,342) node [anchor=north west][inner sep=0.75pt]   [align=left] {HW move};
\draw (76,470.4) node [anchor=north west][inner sep=0.75pt]    {$NS_{5}$};
\draw (182,470.4) node [anchor=north west][inner sep=0.75pt]    {$NS_{5}$};
\draw (272,470.4) node [anchor=north west][inner sep=0.75pt]    {$NS_{5}$};
\draw (366,470.4) node [anchor=north west][inner sep=0.75pt]    {$NS_{5}$};
\draw (401,400.4) node [anchor=north west][inner sep=0.75pt]    {$D_{3}$};
\draw (136,360.4) node [anchor=north west][inner sep=0.75pt]    {$D_{5}$};
\draw (326,362.4) node [anchor=north west][inner sep=0.75pt]    {$D_{5}$};

\end{tikzpicture}

\end{center}
\caption{\label{fig:braneCon} Brane moves to get the mirror pairs in figure \ref{fig:U12r}. Upper: the upper theory in figure \ref{fig:U12r} is the effective theory on the $D_3$ branes intersecting several $(1,1)$-branes and $NS_5$ branes \cite{Kitao:1998mf}. Middle: $\tau\rightarrow \frac{\tau}{1-\tau}$ action sends $(1,1)$-branes to $D_5$ branes while leaves $NS_5$ branes unchanged. Lower: one further moves $D_5$ branes \cite{Hanany:1996ie} to get the lower theory in figure \ref{fig:U12r}.  }
\end{figure}

The mirror pair in figure \ref{fig:U12r} can also be understood from brane moves  shown in figure \ref{fig:braneCon}. This construction is the generalization of the mirror pairs discussed in \cite{Hosomichi:2008jd} to linear quivers.

\section{$\CN=4$ CSM theories coupled with $T_2$ theories}
\label{sec:indCBT2}

\begin{figure}
\begin{center}

\tikzset{every picture/.style={line width=0.75pt}} 

\begin{tikzpicture}[x=0.75pt,y=0.75pt,yscale=-1,xscale=1]

\draw   (99,38) .. controls (99,18.67) and (114.67,3) .. (134,3) .. controls (153.33,3) and (169,18.67) .. (169,38) .. controls (169,57.33) and (153.33,73) .. (134,73) .. controls (114.67,73) and (99,57.33) .. (99,38) -- cycle ;
\draw   (3,212) .. controls (3,192.67) and (18.67,177) .. (38,177) .. controls (57.33,177) and (73,192.67) .. (73,212) .. controls (73,231.33) and (57.33,247) .. (38,247) .. controls (18.67,247) and (3,231.33) .. (3,212) -- cycle ;
\draw    (134,123) -- (134,73) ;
\draw   (133.88,123) -- (169.75,164) -- (98,164) -- cycle ;
\draw   (193,212) .. controls (193,192.67) and (208.67,177) .. (228,177) .. controls (247.33,177) and (263,192.67) .. (263,212) .. controls (263,231.33) and (247.33,247) .. (228,247) .. controls (208.67,247) and (193,231.33) .. (193,212) -- cycle ;
\draw    (98,164) -- (67,191) ;
\draw    (170,164) -- (202,189) ;

\draw (124,139.4) node [anchor=north west][inner sep=0.75pt]    {$T_{N}$};
\draw (104,29.4) node [anchor=north west][inner sep=0.75pt]    {$SU( N)_{k_1}$};
\draw (5,201.4) node [anchor=north west][inner sep=0.75pt]    {$SU( N)_{k_2}$};
\draw (197,202.4) node [anchor=north west][inner sep=0.75pt]    {$SU( N)_{k_3}$};

\end{tikzpicture}

\end{center}
\caption{\label{fig:CSoneTN}CS theory coupled to a single $T_N$ theory.}
\end{figure}
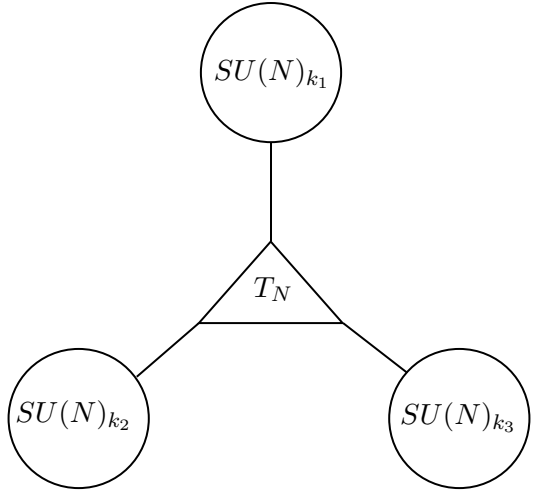

We now shift our attention to CS theories couple to (3d compactification of) $T_N$ theories (see figure \ref{fig:CSoneTN} for an example). The $T_N$ theory is the basic building block of 4d class-S theories with flavor group $SU(N)^3$, and it was argued in \cite{Assel:2022row} that after compactifying to 3d, one can couple each flavor $SU(N)$ with CS term with levels $k_1$, $k_2$ and $k_3$. The resulting 3d theory enhances to $\CN=4$ in the IR if
\begin{equation}
\label{eq:TNbalance}
\frac{1}{k_1}+\frac{1}{k_2}+\frac{1}{k_3}=0.
\end{equation}
One can also take multiple $T_N$ theories and glue them to get more complicated $\CN=4$ theories like what we did in bifundamental cases.
We will  mainly focus on cases when $N=2$ as  $T_2$ is free theory.

\subsection{$\CN=4$ CSM theories with one $T_2$ theory}

Consider only one $T_2$ theory, its  index is
\begin{equation}
\begin{split}
\CI^{T_2,(k_1,k_2,k_3)}
=&\sum_{m_1,m_2,m_3\geq0}\int\left(\prod_{i=1}^3\frac{1}{W(m_i)}\frac{dz_i}{2\pi iz_i}z_i^{2k_im_i}Z_{vec}(\{z_i,m_i\})\right)Z_{T_2}(\{z_i,m_i\}),
\end{split}
\end{equation}
where $Z_{T_2}(\{z_i,m_i\})$ is the index of the trifundamental hyper
\begin{equation}
\begin{split}
&Z_{T_2}(\{z_i,m_i\})\\
=&\prod_{i,j,k=\pm 1}\left(\frac{x}{t^2}\right)^{\frac{1}{2}|im_1+jm_2+km_3|}
\frac{ ( (-1)^{im_1+jm_2+km_3} t^{-1} x^{\frac{3}{2}+ |im_1+jm_2+km_3|} z_1^i z_2^j z_3^k ; x^2)_\infty }{( (-1)^{im_1+jm_2+km_3} t x^{\frac{3}{2}+ |im_1+jm_2+km_3|} z_1^i z_2^j z_3^k ; x^2)_\infty}.
\end{split}
\end{equation}
In the computation we just set  fugacities of any discrete symmetry to be $1$.

\subsubsection{Examples of the full index}

First we consider  $\frac{k_1}{2}=\frac{k_2}{2}=-k_3=k>0$ so the balancing condition \ref{eq:TNbalance} is satisfied.  For the theory to be non-bad, $k$ has to be greater or equal to $3$. The normalized single letter indices for the first few values of $k$ are
\begin{equation}
\begin{split}
\tilde{\CI}^{T_2,(6,6,-3)}
=&(2t^4-1)x^2+(t^6-t^2+t^{-2})x^3-(t^4+1)x^4 \\
&+(t^6-3t^2)x^5+(-t^{12}+t^8+3)x^6+\cdots,\\
\tilde{\CI}^{T_2,(8,8,-4)}
=&(t^4-1)x^2+t^{-2}x^3+(t^8-1)x^4 +(t^{10}-t^6-t^2)x^5+(-t^8+t^4+1)x^6\\
&+(t^{10}-2t^6-t^2)x^7+(-t^8+t^4-1)x^8+\cdots, \\
\tilde{\CI}^{T_2,(10,10,-5)}
=&(t^4-1)x^2+t^{-2}x^3-x^4 -t^2x^5+(t^{12}+t^4+1)x^6\\
&+(t^{14}-t^{10}-t^2)x^7+(-t^{12}-t^4-1)x^8+\cdots,
\end{split}
\end{equation}
The Coulomb limits of these indices are $1$ and the Higgs indices have closed form expression which will discussed later. Our results are the same as results in \cite{Comi:2023lfm} up to the known order.

Another set of balancing $k_i$'s are $\frac{k_1}{6}=\frac{k_2}{3}=-\frac{k_3}{2}=k>0$. For example, when $(k_1,k_2,k_3)=(12,6,-4)$, the indices is
\begin{equation}
\begin{split}
\tilde{\CI}^{T_2,(12,6,-4)}
=&(t^4-1)x^2+t^{-2}x^3-x^4 -t^2x^5+(t^{12}+t^4+1)x^6\\
&+(t^{14}-t^{10}-t^2)x^7+(-t^{12}-t^4-1)x^8+\cdots.
\end{split}
\end{equation}


\subsubsection{The Higgs branch}

To study the Higgs index, again we first write down the index of a $V_{m_1,m_2,m_3}$ monopole in terms of $q$ and $\tilde{q}$,
\begin{equation}
\left(\prod_{i=1}^3z_i^{2k_il_i}\right)q^{-\sum_{i=1}^3|m_i|} \tilde{q}^{\frac{1}{4}\sum_{i,j,k=\pm 1}|im_1+jm_2+km_3| - \sum_{i=1}^3|m_i|}.
\end{equation}
Because $\frac{1}{4}\sum_{i,j,k=\pm 1}|im_1+jm_2+km_3| - \sum_{i=1}^3|m_i|\geq0$,  one can take the Higgs limit ($\tilde{q}\rightarrow 0$) before integration. Denote by $d$ the least common multiplier  $\mathrm{lcm}(|k_1|,|k_2|,|k_3|)$
 of $k_1$, $k_2$, and $k_3$, then the Higgs index receives contribution only from monopoles with magnetic charge $n(l_1,l_2,l_3)\equiv n(|d/k_1|,|d/k_2|,|d/k_3|)$ with $n$ being a non-negative integer. The full Higgs index is then
\begin{equation}
\label{eq:HindexT2}
\CI^{T_2,(k_1,k_2,k_3)}_{H}
=\frac{1-q^{2D(k_1,k_2,k_3)}}{(1-q^2)(1-q^{D(k_1,k_2,k_3)-1})(1-q^{D(k_1,k_2,k_3)})},
\end{equation}
where $D(k_1,k_2,k_3)=d-l_1-l_2-l_3+1=\mathrm{lcm}(|k_1|,|k_2|,|k_3|)\left(1-\frac{1}{|k_1|}-\frac{1}{|k_2|}-\frac{1}{|k_3|}\right)+1$. Interestingly, the Higgs index is the same as the Hilbert series of the affine variety $\bbC^2 / \hat{D}_{D(k_1,k_2,k_3) + 1} $ with the defining equation \cite{Benvenuti:2006qr}
\begin{equation}
u^2+v^2w=w^{ D(k_1,k_2,k_3) }.
\end{equation}
 Here $\hat{D}_{n+2} \subset SU(2)$ is the binary dihedral group generated by
\begin{equation}
\left(
\begin{array}{cc}
\omega_{2n} & 0 \\ 0 & \omega_{2n}^{-1}
\end{array}
\right), \quad 
\left(
\begin{array}{cc}
0 & i \\ i & 0
\end{array}
\right),
\end{equation}
with $\omega_{2n}\equiv e^{\pi i /n}$ being the $2n$-th root of unity. From the Higgs index, we conjecture that the Higgs branch of the CS-$T_2$ theory is $\bbC^2 / \hat{D}_{D(k_1,k_2,k_3) + 1}$
\begin{equation}
\CM_H^{T_2(k_1,k_2,k_3)} = \bbC^2 / \hat{D}_{D(k_1,k_2,k_3) + 1},
\end{equation}
which is also observed in \cite{Comi:2023lfm}.

Notice that the Higgs index only depends on $D(k_1,k_2,k_3)$, therefore theories with the same value of $D(k_1,k_2,k_3)$ have the same Higgs branch although the CS levels can be different. For example, $(k_1,k_2,k_3)=(8,8,-4)$ or $(12,4,-3)$ both have $D=5$, so 
\begin{equation}
\CI^{T_2,(8,8,-4)}_H=\CI^{T_2,(12,4,-3)}_H=\frac{1-q^{10}}{(1-q^2)(1-q^4)(1-q^5)}.
\end{equation}
Another example is $(k_1,k_2,k_3)=(10,10,-5)$ or $(12,6,-4)$ which have $D=7$, so
\begin{equation}
\CI^{T_2,(10,10,-5)}_H=\CI^{T_2,(12,6,-4)}_H=\frac{1-q^{14}}{(1-q^2)(1-q^6)(1-q^{7})}.
\end{equation}
However, their full indices are not the same, so although these theories have the same Higgs and Coulomb index, they are not dual to each other.
 \subsection{$\CN=4$ CS theory coupled to two $T_2$'s}

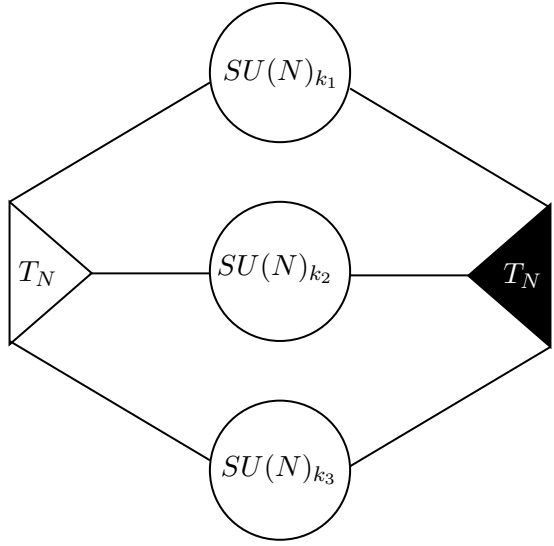
\begin{figure}
\begin{center}

\tikzset{every picture/.style={line width=0.75pt}} 

\begin{tikzpicture}[x=0.75pt,y=0.75pt,yscale=-1,xscale=1]

\draw   (210,35) .. controls (210,15.67) and (225.67,0) .. (245,0) .. controls (264.33,0) and (280,15.67) .. (280,35) .. controls (280,54.33) and (264.33,70) .. (245,70) .. controls (225.67,70) and (210,54.33) .. (210,35) -- cycle ;
\draw   (210,135) .. controls (210,115.67) and (225.67,100) .. (245,100) .. controls (264.33,100) and (280,115.67) .. (280,135) .. controls (280,154.33) and (264.33,170) .. (245,170) .. controls (225.67,170) and (210,154.33) .. (210,135) -- cycle ;
\draw   (151,135.88) -- (110,171.75) -- (110,100) -- cycle ;
\draw    (110,100) -- (210,40) ;
\draw    (110,170) -- (210,230) ;
\draw    (150,136) -- (210,136) ;

\draw   (210,235) .. controls (210,215.67) and (225.67,200) .. (245,200) .. controls (264.33,200) and (280,215.67) .. (280,235) .. controls (280,254.33) and (264.33,270) .. (245,270) .. controls (225.67,270) and (210,254.33) .. (210,235) -- cycle ;
\draw  [fill={rgb, 255:red, 0; green, 0; blue, 0 }  ,fill opacity=1 ] (339,137.12) -- (380,101.25) -- (380,173) -- cycle ;
\draw    (380,173) -- (280,233) ;
\draw    (380,103) -- (280,43) ;
\draw    (340,137) -- (280,137) ;

\draw (113,128.4) node [anchor=north west][inner sep=0.75pt]    {$T_{N}$};
\draw (215,26.4) node [anchor=north west][inner sep=0.75pt]    {$SU( N)_{k_1}$};
\draw (212,124.4) node [anchor=north west][inner sep=0.75pt]    {$SU( N)_{k_2}$};
\draw (214,225.4) node [anchor=north west][inner sep=0.75pt]    {$SU( N)_{k_3}$};
\draw (355,129.4) node [anchor=north west][inner sep=0.75pt]  [color={rgb, 255:red, 255; green, 255; blue, 255 }  ,opacity=1 ]  {$T_{N}$};

\end{tikzpicture}

\end{center}
\caption{\label{fig:CS2TN}CS theory coupled to two $T_N$ theories. Black nodes means $T_N$ theory with opposite $R_H-R_C$ charges as the white one.}
\end{figure}

Now we consider CS theories coupled to two $T_N$ theories as shown in figure \ref{fig:CS2TN}. Again, in order to enhance to $\CN=4$, the CS levels should satisfy the balancing condition $1/k_1+1/k_2+1/k_3=0$, moreover, these two $T_N$ theory should have opposite charges under $R_H-R_C$ \cite{Assel:2022row}. All these theories should be self-mirror by construction and its index is
\begin{equation}
\begin{split}
\CI^{T_2,(k_1,k_2,k_3)}
=&\sum_{m_1,m_2,m_3\geq0}\int\left(\prod_{i=1}^3\frac{1}{W(m_i)}\frac{dz_i}{2\pi iz_i}z_i^{2k_im_i}Z_{vec}(\{z_i,m_i\})\right)\\
&~~~~~~~~~~~~~~~~~~\times Z_{T_2}(\{z_i,m_i\};t) Z_{T_2}(\{z_i,m_i\}; t^{-1}).
\end{split} 
\end{equation}

For $N=2$, $(k_1,k_2,k_3)=(2k,2k,-k)$, we compute an example when $(k_1,k_2,k_3)=(4,4,-2)$ or $(6,6,-3)$,
\begin{equation}
\begin{split}
\tilde{\CI}^{2T_2,(4,4,-2)}
=&x+(t^4+1+t^{-4})x^2-(t^4+2t^2+3+2t^{-2}+t^{-4})x^3\\
&+(t^8+t^6-t^4+3t^2+3t^{-2}-t^{-4}+t^{-6}+t^{-8})x^4\\
&+(t^{10}+t^6+5t^4+8t^2+7+8t^{-2}+5t^{-4}+t^{-6}+t^{-10})x^5+\cdots,\\
\tilde{\CI}^{2T_2,(6,6,-3)}
=&x+(t^4+1+t^{-4})x^2-(t^4+2t^2+3+2t^{-2}+t^{-4})x^3\\
&+(-2t^4+2t^2-1+2t^{-2}-2t^{-4})x^4+\cdots.
\end{split}
\end{equation}
Our results are the same as results of \cite{Comi:2023lfm}.
Apparently these indices are symmetric under $t\rightarrow t^{-1}$, confirming the fact that these theories are self-mirror. Both Higgs and Coulomb indices are
\begin{equation}
\CI^{2T_2,(2k,2k,-k)}_H=\CI^{2T_2,(2k,2k,-k)}_C
=\frac{ 1-q^{4k+2} } {(1-q^2)(1-q^{2k})(1-q^{2k+1}) }.
\end{equation}
Since the twisted trifundamental also modifies the $\Delta-R_C$ charges of the monopole operators so the Higgs branch is modified comparing to the single $T_2$ case. In general the Higgs and Coulomb index of two $T_2$ theories with CS levels $k_1,k_2,k_3$ should be
\begin{equation}
\CI^{2T_2,(k_1,k_2,k_3)}_H=\CI^{2T_2,(k_1,k_2,k_3)}_C =\CI^{2T_2,(2k,2k,-k)}_C
=\frac{ 1-q^{2L+2} } {(1-q^2)(1-q^{L})(1-q^{L+1}) },
\end{equation}
where $L(k_1,k_2,k_3)=\mathrm{lcm}(k_1,k_2,k_3)$.
The moduli space is conjectured to be
\begin{equation}
\CM_H^{2T_2,(k_1,k_2,k_3)}=\CM_C^{2T_2,(k_1,k_2,k_3)}=\bbC^2/\hat{D}_{L(k_1,k_2,k_3)+2}.
\end{equation}
Our results on Higgs branches also matches with  \cite{Comi:2023lfm}.

\subsection{$\CN=4$ CS theory coupled to a web of $T_2$'s}

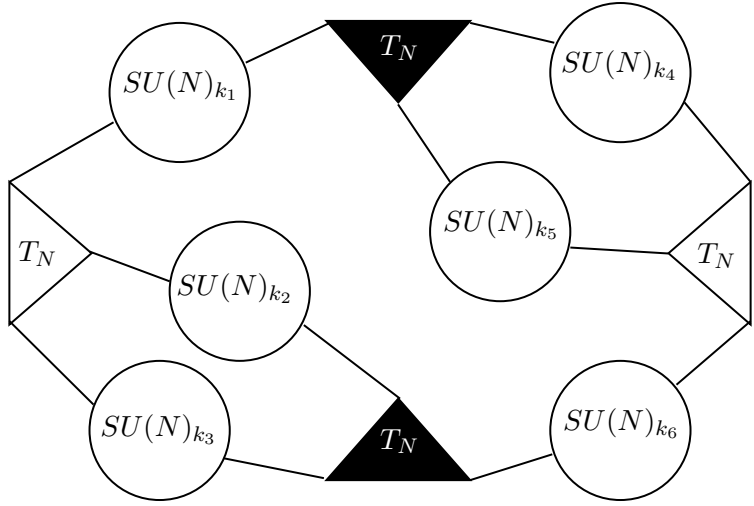
\begin{figure}
\begin{center}

\tikzset{every picture/.style={line width=0.75pt}} 

\begin{tikzpicture}[x=0.75pt,y=0.75pt,yscale=-1,xscale=1]

\draw   (70,55) .. controls (70,35.67) and (85.67,20) .. (105,20) .. controls (124.33,20) and (140,35.67) .. (140,55) .. controls (140,74.33) and (124.33,90) .. (105,90) .. controls (85.67,90) and (70,74.33) .. (70,55) -- cycle ;
\draw   (100,155) .. controls (100,135.67) and (115.67,120) .. (135,120) .. controls (154.33,120) and (170,135.67) .. (170,155) .. controls (170,174.33) and (154.33,190) .. (135,190) .. controls (115.67,190) and (100,174.33) .. (100,155) -- cycle ;
\draw   (61,135.88) -- (20,171.75) -- (20,100) -- cycle ;
\draw   (60,225) .. controls (60,205.67) and (75.67,190) .. (95,190) .. controls (114.33,190) and (130,205.67) .. (130,225) .. controls (130,244.33) and (114.33,260) .. (95,260) .. controls (75.67,260) and (60,244.33) .. (60,225) -- cycle ;
\draw  [fill={rgb, 255:red, 0; green, 0; blue, 0 }  ,fill opacity=1 ] (214.13,60) -- (178.25,19) -- (250,19) -- cycle ;
\draw  [fill={rgb, 255:red, 0; green, 0; blue, 0 }  ,fill opacity=1 ] (214.13,209) -- (250,250) -- (178.25,250) -- cycle ;
\draw   (349,135.88) -- (390,100) -- (390,171.75) -- cycle ;
\draw    (20,100) -- (72,70) ;
\draw    (138,40) -- (180,20) ;
\draw    (60,135) -- (100,150) ;
\draw    (167,172) -- (215,209) ;
\draw    (20,170) -- (62,212) ;
\draw    (127,239) -- (177,249) ;
\draw   (290,45) .. controls (290,25.67) and (305.67,10) .. (325,10) .. controls (344.33,10) and (360,25.67) .. (360,45) .. controls (360,64.33) and (344.33,80) .. (325,80) .. controls (305.67,80) and (290,64.33) .. (290,45) -- cycle ;
\draw   (230,125) .. controls (230,105.67) and (245.67,90) .. (265,90) .. controls (284.33,90) and (300,105.67) .. (300,125) .. controls (300,144.33) and (284.33,160) .. (265,160) .. controls (245.67,160) and (230,144.33) .. (230,125) -- cycle ;
\draw   (290,225) .. controls (290,205.67) and (305.67,190) .. (325,190) .. controls (344.33,190) and (360,205.67) .. (360,225) .. controls (360,244.33) and (344.33,260) .. (325,260) .. controls (305.67,260) and (290,244.33) .. (290,225) -- cycle ;
\draw    (250,20) -- (292,30) ;
\draw    (357,60) -- (390,100) ;
\draw    (214,61) -- (240,100) ;
\draw    (300,133) -- (349,136) ;
\draw    (250,250) -- (291,237) ;
\draw    (390,170) -- (353,202) ;

\draw (23,128.4) node [anchor=north west][inner sep=0.75pt]    {$T_{N}$};
\draw (75,42.4) node [anchor=north west][inner sep=0.75pt]    {$SU( N)_{k_1}$};
\draw (102,144.4) node [anchor=north west][inner sep=0.75pt]    {$SU( N)_{k_2}$};
\draw (64,215.4) node [anchor=north west][inner sep=0.75pt]    {$SU( N)_{k_3}$};
\draw (203.63,24.78) node [anchor=north west][inner sep=0.75pt]  [color={rgb, 255:red, 255; green, 255; blue, 255 }  ,opacity=1 ]  {$T_{N}$};
\draw (203.63,222.78) node [anchor=north west][inner sep=0.75pt]  [color={rgb, 255:red, 255; green, 255; blue, 255 }  ,opacity=1 ]  {$T_{N}$};
\draw (362,129.4) node [anchor=north west][inner sep=0.75pt]    {$T_{N}$};
\draw (294,32.4) node [anchor=north west][inner sep=0.75pt]    {$SU( N)_{k_4}$};
\draw (235,112.4) node [anchor=north west][inner sep=0.75pt]    {$SU( N)_{k_5}$};
\draw (295,212.4) node [anchor=north west][inner sep=0.75pt]    {$SU( N)_{k_6}$};

\end{tikzpicture}

\end{center}
\caption{\label{fig:CS4TN}CS theory with for $T_N$ matters. White/Black nodes indicating opposite $R_H-R_C$ charges.}
\end{figure}

Consider for example the quiver in figure \ref{fig:CS4TN}. Let $N=2$ and $(k_1,k_2,\cdots,k_6)=(-2,4,4,4,4,-2)$. In this particular setting, the theory is enhanced to $\CN=4$, self-mirror and its normalized single letter index is
\begin{equation}
\begin{split}
\tilde{\CI}^{4T_2}=&(2t^4+3+2t^{-4})x^2-(4t^2+4t^{-2})x^3+(-6t^4-9-6t^{-4})x^4\\
&+(4t^6+22t^2+22t^{-2}+4t^{-6})x^5+(2t^{12}+t^8+10t^4+14+10t^{-4}+t^{-8}+2t^{-12})x^6+\cdots.
\end{split}
\end{equation}

If one is interested in Higgs/Coulomb indices only, one can also work out a gluing formula similar to the ones in \ref{sec:linearQuiverInd}.

\subsection{Comments on indices of CS theories coupled to $T_N$}

In this section, we discuss the general formula to compute the index of CS theories couple to $T_N$ or AD matters. For a web of $T_N$ theories, the index in general takes the following schematic form
\begin{equation}
\begin{split}
\label{eq:CS-Tindex}
\CI^{CS-T}=\sum_{\bfm^{(i)}}\int\frac{d\bfz^{(i)}}{2\pi i\bfz^{(i)}}&\prod_i Z_{CS}(\{z^{(i)},m^{(i)}\})Z_{vec}(\{z^{(i)},m^{(i)}\})\\
&\times \prod_{(i_1,i_2,i_3)}Z_{T^+}(\{z^{(i_1)},m^{(i_1)}\},\{z^{(i_2)},m^{(i_2)}\},\{z^{(i_3)},m^{(i_3)}\};t)\\
&\times \prod_{(j_1,j_2,j_3)}Z_{T^-}(\{z^{(j_1)},m^{(j_1)}\},\{z^{(j_2)},m^{(j_2)}\},\{z^{(j_3)},m^{(j_3)}\};t^{-1}).
\end{split}
\end{equation}
Here $T^{\pm}$  represent $T_N$ theories with opposite $R_H-R_C$ charges respectively.

It is now important to compute the exact form of the indices of $T_N$ theories and AD matters, which could be computed using the mirror quiver theory and the relation
\begin{equation}
\CI(x,t)=\CI^{mirror}(x,t^{-1}),
\end{equation}
and symmetries on the Higgs/Coulomb branch of the original theory becomes symmetries on the Coulomb/Higgs branch of the mirror theory.
One can then use the building blocks in \ref{sec:3dindex} to compute the indices of mirror quivers while keeping the fugacities and fluxes of all topological symmetries. Once the $\CN=4$ indices of 3d mirror of $T_N$ theories or AD theories are known, they can be plugged in \ref{eq:CS-Tindex}  to get the $\CN=4$ indices of the corresponding CSM theories. However, as mentioned in \cite{Comi:2023lfm}, this computation is quite complicated and we hope to figure out better ways in the future.

\begin{figure}
\begin{center}

\tikzset{every picture/.style={line width=0.75pt}} 

\begin{tikzpicture}[x=0.75pt,y=0.75pt,yscale=-0.5,xscale=0.5]

\draw   (250,240) .. controls (233.43,240) and (220,226.57) .. (220,210) .. controls (220,193.43) and (233.43,180) .. (250,180) .. controls (266.57,180) and (280,193.43) .. (280,210) .. controls (280,226.57) and (266.57,240) .. (250,240) -- cycle ;
\draw   (170,240) .. controls (153.43,240) and (140,226.57) .. (140,210) .. controls (140,193.43) and (153.43,180) .. (170,180) .. controls (186.57,180) and (200,193.43) .. (200,210) .. controls (200,226.57) and (186.57,240) .. (170,240) -- cycle ;
\draw   (310,300) .. controls (293.43,300) and (280,286.57) .. (280,270) .. controls (280,253.43) and (293.43,240) .. (310,240) .. controls (326.57,240) and (340,253.43) .. (340,270) .. controls (340,286.57) and (326.57,300) .. (310,300) -- cycle ;
\draw   (410,400) .. controls (393.43,400) and (380,386.57) .. (380,370) .. controls (380,353.43) and (393.43,340) .. (410,340) .. controls (426.57,340) and (440,353.43) .. (440,370) .. controls (440,386.57) and (426.57,400) .. (410,400) -- cycle ;
\draw    (200,210) -- (220,210) ;
\draw    (271,231) -- (289,249) ;
\draw    (331,291) -- (349,309) ;
\draw    (371,331) -- (389,349) ;
\draw   (309,120) .. controls (292.43,120) and (279,133.43) .. (279,150) .. controls (279,166.57) and (292.43,180) .. (309,180) .. controls (325.57,180) and (339,166.57) .. (339,150) .. controls (339,133.43) and (325.57,120) .. (309,120) -- cycle ;
\draw   (410,20) .. controls (393.43,20) and (380,33.43) .. (380,50) .. controls (380,66.57) and (393.43,80) .. (410,80) .. controls (426.57,80) and (440,66.57) .. (440,50) .. controls (440,33.43) and (426.57,20) .. (410,20) -- cycle ;
\draw    (270,189) -- (288,171) ;
\draw    (330,129) -- (348,111) ;
\draw    (371,89) -- (389,71) ;
\draw   (250,235) .. controls (236.19,235) and (225,223.81) .. (225,210) .. controls (225,196.19) and (236.19,185) .. (250,185) .. controls (263.81,185) and (275,196.19) .. (275,210) .. controls (275,223.81) and (263.81,235) .. (250,235) -- cycle ;
\draw [line width=1.5]  [dash pattern={on 1.69pt off 2.76pt}]  (352,312) -- (370,330) ;
\draw [line width=1.5]  [dash pattern={on 1.69pt off 2.76pt}]  (350,109) -- (368,91) ;
\draw   (50,240) .. controls (33.43,240) and (20,226.57) .. (20,210) .. controls (20,193.43) and (33.43,180) .. (50,180) .. controls (66.57,180) and (80,193.43) .. (80,210) .. controls (80,226.57) and (66.57,240) .. (50,240) -- cycle ;
\draw    (80,210) -- (100,210) ;
\draw    (120,210) -- (140,210) ;
\draw [line width=1.5]  [dash pattern={on 1.69pt off 2.76pt}]  (100,210) -- (120,210) ;

\draw (237,200) node [anchor=north west][inner sep=0.75pt]  [font=\scriptsize]  {$N$};
\draw (280,142.4) node [anchor=north west][inner sep=0.75pt] [font=\tiny]    {$N-1$};
\draw (280,262.4) node [anchor=north west][inner sep=0.75pt]  [font=\tiny]   {$N-1$};
\draw (42,200) node [anchor=north west][inner sep=0.75pt]  [font=\scriptsize]   {$1$};
\draw (401,362.4) node [anchor=north west][inner sep=0.75pt]  [font=\scriptsize]   {$1$};
\draw (140,200) node [anchor=north west][inner sep=0.75pt]  [font=\tiny]   {$N-1$};
\draw (401,42.4) node [anchor=north west][inner sep=0.75pt] [font=\scriptsize]    {$1$};

\end{tikzpicture}
\end{center}
\caption{\label{fig:mirrorTN}The 3d mirror quiver of $T_N$ theory. All nodes represent a $U(l)$ $\CN=4$ vector multiplet except the middle one which is $SU(N)$. All lines represent bi-fundamental hypermultiplets.}
\end{figure}
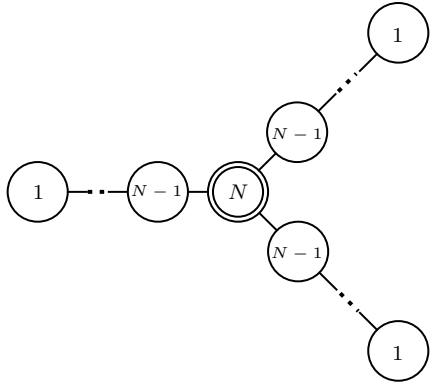

\section{Conclusion}

In this work we computed the indices of several class of $\mathcal{N}=4$ CSM theories, and used them to extract informations of the Higgs/Coulomb branches. In particular, we also provide closed form expressions and gluing formula for Higgs/Coulomb indices. 
The moduli space of vacua of these CSM theories are usually quite special, and usually have quite different structure from the usual $\mathcal{N}=4$ gauge theory, and certainly one should further study those theories.

One of the major challenge is to compute the index of CSM theories built from $T_N$ theory and other type of theories constructed from 
 4d $\mathcal{N}=2$ theories. The theory space of $\mathcal{N}=4$ CSM theory is much larger now and we hope to learn more features of this class of theories.

\acknowledgments

BL and WY are supported by Yau Mathematical Science Center at Tsinghua
University. DX and WY are supported by  national key research
and development program of China (NO. 2020YFA0713000), and NNSF of China with
Grant NO: 11847301 and 12047502. 

\appendix

\section{Review of 3d supersymmetric index}
\label{sec:3dindex}

\subsection{3d $\CN=2$ index}

In this section we first briefly review the 3d $\CN=2$ supersymmetric index (index) which is the supersymmetric partition function on $S^2\times S^1$ \cite{Bhattacharya:2008zy, Bhattacharya:2008bja, Kim:2009wb,Imamura:2011su,Kapustin:2011jm,Dimofte:2011py}. Following the notation of \cite{Aharony:2013dha,Aharony:2013kma,Garozzo:2019ejm}, it is the following trace over states on $S^2\times\bbR$,
\begin{equation}
\CI^{\CN=2}(x,\mathbf{\mu})=\mathrm{Tr}_{\CH}\left[(-1)^{2J_3}x^{\Delta+J_3}\prod_i\mu_i^{T_i}\right],
\end{equation}
where $\Delta$ is the energy times the radius of $S^2$ ($\Delta$ is the conformal dimension for a superconformal theory), $J_3$ is the Cartan generator of the $SO(3)$ isometry of $S^2$, and $T_i$ are charges under non-$R$ global symmetries. The index receives contributions only from BPS states satisfying
\begin{equation}
\Delta-R-J_3=0,
\end{equation}
where $R$ is the $R$-charge.

Localisation leads to the following compact form of the index
\begin{equation}
\CI^{\CN=2}(\{\bfw,\bfn\};x)=\sum_{\bfm}\frac{1}{|W_{\bfm}|}\int\frac{d\bfz}{2\pi i\bfz}Z_{CS}Z_{vec} Z_{mat}.
\end{equation}
Here $\bfz$ is the collective notation for fugacities parametrizing the maximal torus of the gauge group $G$, and $\bfm$ the corresponding GNO magnetic fluxes on $S^2$ which lives in the coweight lattice $P^\vee_G$ of the gauge group $G$.  $W_\bfm$ is the order of the Weyl group of the residual gauge symmetry in the monopole background with fluxes $\bfm$. $\{\bfw,\bfn\}$ are collective notation for possible fugacities and fluxes for the background vector multiplets associated with global symmetries respectively. $Z_{CS}$, $Z_{vec}$ and $Z_{mat}$ are contributions from CS action, 1-loop corrections of vector multiplets and  chiral multiplets respectively. Each term can be expressed explicitly.
\begin{itemize}
\item $Z_{CS}$: The classical contribution from CS action. Denoting by $G$ the gauge group, $k$ the CS level, $Z_{CS}$ is of the form
\begin{equation}
Z_{CS}=\prod_{i=1}^{\mathrm{rank}G} w^{m_i}z_i^{km_i+n},
\end{equation}
with $\{\bfz,\bfm\}$ fugacity and flux of the gauge symmetry, $\{w,n\}$ fugacity and flux of the topological symmetry.
\item $Z_{vec}$: The contribution from $\CN=2$ vector multiplet of the gauge symmetry algebra $\fkg$ is of the form,
\begin{equation}
\label{eq:defN2vec}
Z_{vec}=\prod_{\alpha\in\Delta}x^{-\frac{|\alpha(\bfm)|}{2}}\left(1-(-1)^{\alpha(\bfm)}\bfz^\alpha x^{|\alpha(\bfm)|}\right).
\end{equation}
Here $\Delta$ is the root system of $\fkg$, $\{\bfz,\bfm\}$ fugacity and flux of the gauge symmetry.
\item $Z_{mat}$: The contribution from $\CN=2$ chiral multiplets $X$ transforming in the representation $(R_G, R_f)$ of the gauge and flavor group is
\begin{equation}
\label{eq:defN2mat}
\begin{split}
Z_X=\prod_{\rho\in\mathrm{wt}R_G}\prod_{\sigma\in\mathrm{wt}R_F}&(\bfz^\rho \bfw^{\sigma} x^{r-1})^{-\frac{1}{2}|\rho(\bfm)+\sigma(\bfn)|}\\
&\times\frac{ ((-1)^{\rho(\bfm)+\sigma(\bfn)} \bfz^{-\rho}\bfw^{-\sigma} x^{2-r+|\rho(\bfm)+\sigma(\bfn)|};x^2)_\infty}{( (-1)^{\rho(\bfm)+\sigma(\bfn)} \bfz^{\rho}\bfw^{\sigma} x^{r+|\rho(\bfm)+\sigma(\bfn)|}; x^2)_\infty},
\end{split}
\end{equation}
where $\mathrm{wt}R_G$ and $\mathrm{wt}R_F$ are weights of $R_G$ and $R_F$ respectively. $r$ is the $R$-charge of $X$. $\{\bfz,\bfm\}$ are fugacity and flux of the gauge symmetry, while $\{\bfw,\bfn\}$ fugacity and flux of the background vector multiplets of the flavor symmetry. Finally $(z;q)_\infty$ is the $q$-Pochhammer symbol
\begin{equation}
(z;q)_\infty=\prod_{n=0}^\infty(1-zq^n).
\end{equation}
\end{itemize}

\subsection{3d $\CN=4$ index}

For $\CN=4$ theories, one can first split $\CN=4$ multiplets as $\CN=2$ multiplets and then work out the corresponding index. The $R$-symmetry of the 3d $\CN=4$ theory is $SU(2)_H\times SU(2)_C$. Denoting by $R_H$ and $R_C$ the Cartans of $SU(2)_H\times SU(2)_C$ symmetry, one  chooses the $\CN=2$ subalgebra such that it commutes with the combination $R_H-R_C$, and the $\CN=4$ index can be written as a generalization of the $\CN=2$ index
\begin{equation}
\label{eq:defN4index}
\CI^{\CN=4}(x,t,\mathbf{\mu})=\mathrm{Tr}_{\CH}\left[(-1)^{2J_3}x^{\Delta+J_3}t^{2(R_H-R_C)}\prod_i\mu_i^{T_i}\right].
\end{equation}
And $\mathrm{Tr}_{\CH}$ is summed over BPS states satisfying
\begin{equation}
\Delta - R_{H}-R_{C}-J_3=0.
\end{equation} 
$t$ is the fugacity of $2(R_H-R_C)$, since the 3d mirror symmetry exchanges $SU(2)_H$ and $SU(2)_C$, the mirror symmetry sends $t$ to $t^{-1}$.

For a Lagrangian theory, the index can be computed as the following
\begin{equation}
\label{eq:defIndexN4}
\CI^{\CN=4}(\{\bfw,\bfn\};x,t)=\sum_{\bfm}\frac{1}{|W_{\bfm}|}\int\frac{d\bfz}{2\pi i\bfz}Z_{CS}Z^{\CN=4}_{vec} Z_{hyp}.
\end{equation} 
The CS term $Z_{CS}$ is the same as $\CN=2$ case, while the contributions from $\CN=4$ hypermultiplets and $\CN=4$ vector multiplets are summarized below.
\begin{itemize}
\item $Z_{hyp}$: the contribution of a hypermultiplet $(X,\tilde{X})$ is of the following form.
\begin{equation}
\label{eq:defN4hyp}
\begin{split}
Z_{hyp}=\prod_{\rho\in\mathrm{wt}R_G}\prod_{\sigma\in\mathrm{wt}R_F} &\left(\frac{x}{t^2}\right)^{\frac{1}{2}|\rho(\bfm)+\sigma(\bfn)|}
\frac{ ((-1)^{\rho(\bfm)+\sigma(\bfn)} t^{-1} \bfz^{-\rho}\bfw^{-\sigma} x^{\frac{3}{2}+|\rho(\bfm)+\sigma(\bfn)|};x^2)_\infty}{(  (-1)^{\rho(\bfm)+\sigma(\bfn)} t \bfz^{\rho}\bfw^{\sigma} x^{\frac{1}{2}+|\rho(\bfm)+\sigma(\bfn)|}; x^2)_\infty}\\
&\times\frac{ ((-1)^{\rho(\bfm)+\sigma(\bfn)} t^{-1} \bfz^{\rho}\bfw^{\sigma} x^{\frac{3}{2}+|\rho(\bfm)+\sigma(\bfn)|};x^2)_\infty}{(  (-1)^{\rho(\bfm)+\sigma(\bfn)} t \bfz^{-\rho}\bfw^{-\sigma} x^{\frac{1}{2}+|\rho(\bfm)+\sigma(\bfn)|}; x^2)_\infty}.
\end{split}
\end{equation}
Here $X$ and $\tilde{X}$ are the two $\CN=2$ chirals transforming in the representation $(R_G, R_F)$ and $(\bar{R}_G,\bar{R}_F)$ respectively under the gauge and flavor symmetries. We will also denoting by $(R_G, R_F)$  the representation of the full hypermultiplet $(X,\tilde{X})$.
\item $Z_{vec}^{\CN=4}$: an $\CN=4$ vector multiplet is composed of an $\CN=2$ vector multiplet and an adjoint $\CN=2$  chiral multiplet, so its contribution to the index is
\begin{equation}
\label{eq:defN4vec}
\begin{split}
Z^{\CN=4}_{vec}=&\left(\frac{(t^2x;x^2)_\infty}{ (t^{-2}x;x^2)_\infty}\right)^{\mathrm{rank}\fkg}\\
&\times \prod_{\alpha\in\Delta}\left(\frac{x}{t^2}\right)^{-\frac{|\alpha(\bfm)|}{2}}
\left(1-(-1)^{\alpha(\bfm)}\bfz^\alpha x^{|\alpha(\bfm)|}\right)\frac{ ( (-1)^{\alpha(\bfm)} t^2 \bfz^\alpha  x^{|\alpha(\bfm)|} ; x^2 )_\infty }{ ( (-1)^{\alpha(\bfm)} t^{-2} \bfz^\alpha x^{|\alpha(\bfm)|}; x^2 )_\infty }
\end{split}
\end{equation}
\end{itemize}

\subsection{The Higgs/Coulomb limit of the 3d $\CN=4$ index}

It is also useful to define the following combinations of fugacities
\begin{equation}
q\equiv xt^2,\quad\tilde{q}\equiv x t^{-2},
\end{equation}
and the $\CN=4$ index \ref{eq:defN4index} becomes
\begin{equation}
\begin{split}
\CI^{\CN=4}(x,t,\mathbf{\mu})=&\mathrm{Tr}_{\CH}\left[(-1)^{2J_3}q^{J_3+R_H}\tilde{q}^{J_3-R_C}\prod_i\mu_i^{T_i}\right],\\
=&\mathrm{Tr}_{\CH}\left[(-1)^{2J_3}q^{\Delta-R_C}\tilde{q}^{\Delta-R_H}\prod_i\mu_i^{T_i}\right].
\end{split}
\end{equation}
The second equality uses the BPS condition $\Delta-R_H-R_C-J_3=0$.

The Higgs/Coulomb index is the following limit of the $\CN=4$ index \cite{Razamat:2014pta}:
\begin{itemize}
\item Higgs: $x\rightarrow 0$, $t \rightarrow \infty$ ($\tilde{q}\rightarrow 0$), while $q= xt^2$ fixed,
\item Coulomb: $x\rightarrow 0$, $t \rightarrow 0$ ($q\rightarrow 0$), while $\tilde{q}= xt^{-2}$ fixed.
\end{itemize}
One can show that the Higgs/Coulomb index is the same as the Hilbert series of the chiral ring of the Higgs/Coulomb branch operators of the $\CN=4$ theory when there is no extra superpotential $W$ in the theory. In the following, we will illustrate this point in the Higgs limit and the Coulomb limit is similar.

The gauge invariant chiral operators of a 3d $\CN\geq2$ theory  are 't Hooft monopole operators dressed by matter fields with zero effective mass. Denote a bare chiral monopole operator with magnetic charge $\bfm=(m_1,\cdots,m_r)$ by $V_{\bfm}$. Let $A$ be a gauge or global symmetry of the theory, then the total $A$ charge of the monopole operator $V_{\bfm}$ after quantum correction is
\begin{equation}
Q_A[V_\bfm]=-\sum_jk_{Aj}m_j-\frac{1}{2}\sum_{a=1}^n Q_A[\psi_a]\left|\sum_iQ^a_im_i\right|,
\end{equation}
where $k_{Aj}$ is the CS level between the connection of $A$ and the gauge connection $A_j$, and the sum runs over all fermion matter fields $\psi_a$ and $Q^a_i=Q_i[\psi_a]$ are the gauge electric charges of the fermions. The effective mass of a matter fields $X$ in the monopole back ground is $|\rho_X(\bfm)|$ where $\rho_X$ is the gauge representation of $X$. $V_\bfm$ can only be dressed with positive powers of $X$ with $|\rho_X(\bfm)|=0$.

In supersymmetric index, $Z_{CS}$ and the prefactors $\prod_{\alpha}x^{-\frac{|\alpha(\bfm)|}{2}}$, $\prod_{\rho,\sigma}(\bfz^\rho \bfw^{\sigma} x^{r-1})^{-\frac{1}{2}|\rho(\bfm)+\sigma(\bfn)|}$, $\prod_{\rho,\sigma} \left(\frac{x}{t^2}\right)^{\frac{1}{2}|\rho(\bfm)+\sigma(\bfn)|}$, and $\prod_{\alpha}\left(\frac{x}{t^2}\right)^{-\frac{|\alpha(\bfm)|}{2}}$ in \ref{eq:defN2vec}, \ref{eq:defN2mat}, \ref{eq:defN4hyp} and \ref{eq:defN4vec} keep track of the total charges of $V_\bfm$ of the theory. If $\Delta-R_H\geq0$ for all $V_\bfm$, taking the limit $\tilde{q}\rightarrow 0$ selects the monopole operators in the Higgs branch. For the matter contribution, since $tx^{\frac{1}{2}+|\rho(\bfm)|}=q^{\frac{1}{2}+\frac{1}{2}|\rho(\bfm)|}\tilde{q}^{\frac{1}{2}|\rho(\bfm)|}$ and $x^2=q\tilde{q}$, the limit $\tilde{q}\rightarrow 0$ selects automatically matters with zero effective mass $|\rho(\bfm)|=0$. Therefore, we see that
\begin{equation}
\lim_{\tilde{q}\rightarrow 0} Z_{CS}(\bfm)Z_{vec}(\bfm)Z_{hyp}(\bfm)
\end{equation}
counts $V_{\bfm}$ with all possible dressing of matter fields with zero effective mass. Integration over gauge fugacities keeps gauge invariant operators with $\bfm$. In the end, we sum over all possible magnetic charges $\bfm$. The resulting is in the same spirit as the Hilbert series of chiral operators in a theory without superpotental in \cite{Cremonesi:2016nbo}, therefore is the same as the Hilbert series of the Higgs branch chiral operators. The Higgs or Coulomb index are much simpler than the full index and in many cases one can give closed form expression.


\section{The proof of a technical identity}
\label{app:sec:mirrorId}

In this section we prove an identity used to show that $\CI_H = \CI^{mirror}_C$ in section \ref{sec:mirrorLinearQuiver}. Define
\begin{equation}
A=\frac{1}{(1-q)^{2r-1}} \sum_{\tm_1,\cdots,\tm_r\in\bbZ} q^{|\tm_1|/2}q^{|\tm_1-\tm_2|/2} q^{|\tm_2|/2}q^{|\tm_2-\tm_3|/2}\cdots q^{|\tm_{r-1}-\tm_r|/2}  q^{|\tm_r|/2},
\end{equation}
and 
\begin{equation}
B=\int \prod_{i=1}^{r-1}\frac{dz_i}{2\pi i z_i} \frac{1}{(1-q^{\frac{1}{2}} z_1^\pm)^2} \frac{1}{(1-q^{\frac{1}{2}} (z_2/z_1)^\pm) }\frac{1}{(1-q^{\frac{1}{2}} z_2^\pm)} \frac{1}{(1-q^{\frac{1}{2}} (z_3/z_2)^\pm) }\cdots \frac{1}{(1-q^{\frac{1}{2}} z_{r-1}^\pm)^2},
\end{equation}
we need to show that $A=B$.

Firstly, expanding the $(1-q)^{-2r+1}$ factor in terms of the geometric series, we rewrite $A$ as
\begin{equation}
A=\sum_{(\bfm,\bfn)\in S_A}\prod_{j=1}^{2r-1} q^{\frac{|m_j|}2 + n_j},
\end{equation}
where $\bfm=(m_1,\cdots,m_{2r-1} )\in \bbZ^{2r-1}$ and $\bfn = (n_1,\cdots, n_{2r-1})\in \bbZ_{\geq0}^{2r-1}$, and $S_A$ is the following set
\begin{equation}
S_A=\left\{ (\bfm,\bfn) \in \bbZ^{2r-1}\times \bbZ_{\geq0}^{2r-1} |   m_{2j-1}+m_{2j} - m_{2j+1} =0 ,1\leq j\leq r-1\right\}.
\end{equation}
Basically we redefine $m_{2j-1}=-\tm_j$, $m_{2j}=\tm_j-\tm_{j+1}$, $m_{2j+1}= - \tm_{j+1}$, and then imposes the condition on $m_{2j-1} + m_{2j} - m_{2j+1} = 0$ when summing over $\bfm$.

Secondly, using the geometric series $\frac{1}{(1-q^{\frac{1}{2}} z^\pm)} = \sum_{k,l\geq 0} q^{\frac{k+l}{2}} z^{k-l}$, and the fact that the integrals pick up the coefficient of the constant term ($(z_1z_2\cdots z_{r-1})^0$ term) in the integrand, we can rewrite $B$ as
\begin{equation}
B = \sum_{ ( \bfk,\bfl ) \in S_B } \prod_{i=1}^{2r-1} q^{\frac{k_i+l_i }{2} },
\end{equation}
where $\bfk=\{k_1,k_2,\cdots,k_{2r-1}\}$ and $\bfl=\{l_1,l_2,\cdots,l_{2r-1}\}$ and $S_B$ is the allow set of $(\bfk,\bfl)$ which is
\begin{equation}
S_B=\left\{ (\bfk,\bfl) \in \bbZ_{\geq 0}^{4r-2} |   (k_{2j-1}-l_{2j-1})+(k_{2j}-l_{2j})-(k_{2j+1}-l_{2j+1})=0 ,1\leq j\leq r-1\right\}.
\end{equation}
Here we label $\bfk$ and $\bfl$ such that in the integrand the power of $z_j$ in the sum is $(k_{2j-1}-l_{2j-1})+(k_{2j}-l_{2j})-(k_{2j+1}-l_{2j+1})$, then picking up the constant term in $z_j$ imposes the condition $(k_{2j-1}-l_{2j-1})+(k_{2j}-l_{2j})-(k_{2j+1}-l_{2j+1})=0$.

To prove $A=B$, we then need to show that there is a bijection between $S_A$ and $S_B$ which maps $|m_i|/2 + n_i$ to $(k_i+l_i)/2$. We define the map $f:\bbZ^{4r-2}_{\geq0}\rightarrow \bbZ^{2r-1}\times \bbZ^{2r-1}_{\geq0}$ by
\begin{equation}
m_i = k_i-l_i,\quad n_i=\frac{k_i+l_i-|k_i-l_i|}{2},\quad 1\leq i\leq 2r-1.
\end{equation}
By definition $f$ sends $(k_i+l_i)/2$ into $|m_i|/2+n_i$ and $(k_{2j-1}-l_{2j-1})+(k_{2j}-l_{2j})-(k_{2j+1}-l_{2j+1})=0$ is translated into $m_{2j-1}+m_{2j} - m_{2j+1} =0$. Because $|k_i-l_i|$ is either $k_i-l_i$ or $l_i-k_i$, $n_i$ is either $l_i$ or $k_i$, hence a non-negative integer. Moreover, $f$ is piece-wise linear and invertible when $k_i\geq l_i$ or $k_i\leq l_i$, therefore $f$ is a bijection between $S_A$ and $S_B$. Combining all these facts we have
\begin{equation}
A=B.
\end{equation}

\section{$U(2)_{3}\times U(1)_{-2}$ CSM theory: a non-example}

In this section, we consider the index of a theory with non-balanced levels so that the theory will not enhance to $\CN=4$. Consider $U(2)_3\times U(1)_{-2}$ CSM theory with levels $k_1=3$ and $k_2=-2$. The index (with fugacity of topological symmetry setting to $1$) is
\begin{equation}
\CI^{U(2)_{3}\times U(1)_{-2}}_H
=1+t^2x +(t^4-2)x^2 + (t^6+t^{-2})x^3 +(t^8-2)x^4+(t^{10}-3t^2)x^5+\cdots.
\end{equation}
If we take formally the "Higgs" limit, the result is simply 
\begin{equation}
\frac{1}{1-q},
\end{equation}
which means the "Higgs" branch is $\bbC$. Since the resulting manifold is not hyper-Kahler, the theory can not be $\CN=4$.

In \cite{Evtikhiev:2017heo}, the author discussed necessary conditions on the indices so that the theory may have an enhanced supersymmetry. However, in this example, the condition is not strong enough to rule out the $\CN=4$ susy from the $\CN=2$ index only. The fugacity $t$ is important here for checking the $\CN=4$ supersymmetry.

\bibliographystyle{jhep}
\bibliography{ref.bib}
\end{document}